\documentclass[aps,prd,eqsecnum,11pt,nofootinbib,notitlepage]{revtex4-1}
\usepackage{amsmath}
\usepackage{amsfonts}
\usepackage{amssymb}
\usepackage{graphics}
\usepackage{enumitem}

\IfFileExists{dsfont.sty}
        {\usepackage{dsfont}
         \let\mathbb=\mathds
         \newcommand{\id}{\mathds{1}}}
        {\typeout{Package dsfont.sty was not found, using alternative macros.}
         \let\mathds=\mathbb
         \newcommand{\id}{\mbox{1 \kern-.59em {\rm l}}}}

\usepackage{slashed}
\usepackage{units}
\usepackage{setspace}
\allowdisplaybreaks
\newcommand{\cA}{\mathcal{A}}
\newcommand{\cB}{\mathcal{B}}

\newcommand{\cE}{\mathcal{E}}

\newcommand{\cH}{\mathcal{H}}

\newcommand{\cL}{\mathcal{L}}

\newcommand{\cS}{\mathcal{S}}

\newcommand{\cX}{\mathcal{X}}

\newcommand{\nn}{\nonumber}
\newcommand{\sdfrac}[2]{\mbox{\small$\displaystyle\frac{#1}{#2}$}}

\newcommand{\pa}{\partial}                                              
\renewcommand{\a}{\alpha}
\newcommand{\bet}{\beta}
\newcommand{\g}{\gamma}
\newcommand{\del}{\delta}
\newcommand{\e}{\epsilon}

\newcommand{\z}{\zeta}

\newcommand{\lam}{\lambda}
\newcommand{\m}{\mu}
\newcommand{\n}{\nu}

\newcommand{\s}{\sigma}

\newcommand{\vf}{\varphi}

\newcommand{\w}{\omega}

\newcommand{\G}{\Gamma}
\newcommand{\D}{\Delta}

\newcommand{\La}{\Lambda}
\newcommand{\Y}{\Upsilon}

\newcommand{\W}{\Omega}

\newcommand{\R}{\mathds{R}}

\def\nbox#1#2{\vcenter{\hrule \hbox{\vrule height#2in
\kern#1in \vrule} \hrule}}
\def\sq{\,\raise.5pt\hbox{$\nbox{.09}{.09}$}\,}
\def\sqb{\,\raise.5pt\hbox{$\overline{\nbox{.09}{.09}}$}\,}

\newcommand{\na}{\nabla}
\newcommand{\ra}{\rightarrow}
\newcommand{\bea}{\begin{eqnarray}}
\newcommand{\eea}{\end{eqnarray}}
\newcommand{\be}{\begin{equation}}
\newcommand{\ee}{\end{equation}}
\newcommand{\bes}{\begin{subequations}}
\newcommand{\ees}{\end{subequations}}
\def\lag{\langle}
\def\rag{\rangle}

\def\nn{\nonumber\\}
\newcommand{\bx}{{\bf x}}
\newcommand{\bxp}{{\bf x'}}
\newcommand{\bk}{{\bf k}}
\newcommand{\bn}{{\bf \hat N}}
\newcommand{\br}{{\bf \hat r}}
\newcommand{\bR}{{\bf r}}

\newcommand{\mQ}{{\mathds{Q}}}

\newcommand{\tilt}{\tilde t}

\usepackage{accents}
\newcommand*{\dt}[1]{%
  \accentset{\mbox{\large\bfseries .}}{#1}}
\newcommand*{\ddt}[1]{%
  \accentset{\mbox{\large\bfseries .\hspace{-0.25ex}.}}{#1}}

\usepackage[bookmarksnumbered=true,breaklinks=true]{hyperref}

\allowdisplaybreaks[4]
\begin{document}
\topmargin -1.2cm
\begin{flushright} {LA-UR-16-23649}
\end{flushright}\vspace{1em}
\title{\Large Scalar Gravitational Waves in the Effective Theory of Gravity}
\vspace{1cm}

\author{Emil Mottola}
\affiliation{Theoretical Division, T-2\\
MS B285\\
Los Alamos National Laboratory\\
Los Alamos, NM 87545 USA\vspace{5mm}}
\email{emil@lanl.gov}

\date{\version}

\begin{abstract}
\vspace{1cm}
As a low energy effective field theory, classical General Relativity receives an infrared relevant modification
from the conformal trace anomaly of the energy-momentum tensor of massless, or nearly massless, quantum fields.
The local form of the effective action associated with the trace anomaly is expressed in terms of a dynamical scalar
field that couples to the conformal factor of the spacetime metric, allowing it to propagate over macroscopic distances.
Linearized around flat spacetime, this semi-classical EFT admits scalar gravitational wave solutions in addition
to the transversely polarized tensor waves of the classical Einstein theory. The amplitude of the scalar wave modes, 
as well as their energy and energy flux which are positive and contain a monopole moment, are computed. 
Astrophysical sources for scalar gravitational waves are considered, with the excited gluonic condensates in the 
interiors of neutron stars in merger events with other compact objects likely to provide the strongest burst signals.
\end{abstract}
\date{\today}
\maketitle

\newpage
\setcounter{page}{1}
\section{Introduction: Effective Field Theory and Quantum Anomalies}
\label{Sec:Intro}

Einstein's classical theory of General Relativity (GR) predicts the existence of gravitational waves, created
in significant amounts by collapsing and rotating binary systems \cite{GW}. Almost a century after this prediction,
the Laser Interferometer Gravitational-wave Observatory (LIGO) and LSC collaboration have reported the first
direct detection of gravitational waves from the inspiral and coalescence of two intermediate mass black hole
candidates of approximately $30 M_{\odot}$ each \cite{LIGO}.

As this confirmation of classical GR is added to its already impressive successes in the astrophysical domain,
quantum field theory (QFT) has also achieved a parallel and even more detailed confirmation in the microscopic
domain, culminating in the Standard Model unifying the weak and electromagnetic forces, and the discovery of the
Higgs boson \cite{Higgs}.  At the most microscopic scales probed by experiment, the fundamental
principles of quantum mechanics and special relativistic Lorentz invariance, the twin pillars of QFT,
clearly hold.

These successes of classical GR and QFT, each in their respective domains, require that they be
brought together in some consistent framework, preserving the main features of each, at least
in an approximate sense, and in a quantitatively controllable manner. Absent a unifying synthesis
which is able to completely describe and predict phenomena across the broad spectrum of scales
from microphysics to cosmology, the best techniques presently at our disposal for a partial union of
these disparate theories, with a minimum of additional assumptions, are those of Effective Field Theory (EFT)
and semi-classical background field methods, such as QFT in curved spacetime \cite{BirDav}.

Effective Field Theory (EFT) methods have been developed and successfully applied in a number of areas from nuclear
and particle physics to condensed matter physics \cite{EFT, EFText}. The key hypothesis in the EFT approach is the 
{\it decoupling} of short and long distance scales, making possible a description of physics at energies far below an ultraviolet
(UV) cutoff scale $M$ by a set of local operators consistent with low energy symmetries. The low energy effective action
is constructed by truncating at some finite order the sum of such local terms in increasing numbers of derivatives divided
by powers of $M$, and fixing their coefficients from experiment. The resulting effective theory is expected then to apply
to all physical processes with energy-momenta $k \ll M$, or equivalently, at distance scales $\ell \gg M^{-1}$, with an
error of the first order of terms in $k/M$ which have been neglected in the derivative expansion. This is a systematic
approach, consistent with general QFT principles, relying on the decoupling theorem of ultra-short distance physics
at very large energy scales from the low energy, longer wavelength processes of interest \cite{AppCar}, which applies
to both renormalizable and non-renormalizable effective theories. Indeed every QFT, including also the Standard Model
itself may be regarded as an EFT \cite{EFT,EFText}.

The EFT method can be extended to gravity with no essential difficulty.  Classical GR itself is best understood
as an EFT for energies much less than the Planck energy $M_{Pl}\,c^2 = (\hbar c^5/G)^{\frac{1}{2}}$, or on distance
scales much greater than the Planck length $L_{Pl} = (\hbar G/c^3)^{\frac{1}{2}} \simeq 2 \times 10^{-33}$ cm.
Since the local EFT corrections to the gravitational action involve higher powers of the Riemann curvature tensor and its
contractions, these gravitational corrections to classical GR are highly suppressed in the weak field limit of nearly flat spacetime
or for distances $r \gg L_{Pl}$ \cite{DonBur}. If these local higher order curvature corrections would be the only possible QFT
induced corrections to classical gravity, quantum effects would be entirely negligible and could safely be ignored on macroscopic
distance scales, at least if the Einstein Equivalence Principle and metric foundation of GR are rigorously adhered to.

The semi-classical approach to EFT relies on the other hand upon the development of QFT in background curved spacetimes,
treating the metric classically (at least at first), while the fields propagating on this classical curved background are allowed to be
fully quantum. Since the metric is otherwise arbitrary, QFT in curved spacetime is a particular application of the background field
method \cite{DeW}, not relying upon any perturbative or local expansion around flat space, static sources or weak gravitational fields.
When the quantum fields are completely `integrated out' in an arbitrary curved background and QFT divergences are either cut off at
short distances, or renormalized in the standard way consistent with general covariance, an exact effective action is produced which
contains the higher curvature squared terms characteristic of the previous EFT expansion in local invariants. Hence the semi-classical
background field method and the EFT method of cataloguing local operators are closely related. However, in integrating over the
quantum fields at all scales, the background field method and quantum effective action approach captures one important but somewhat
more subtle consequence of QFT easily overlooked in a more phenomenological Wilsonian EFT expansion in local curvature invariants,
namely the generally {\it non-local} and {\it infrared} effects of quantum anomalies.

The best known and most instructive example of a quantum anomaly is the axial current anomaly in either QED or QCD. At the
classical level the axial current $j^{5\mu}$ is a Noether current whose divergence is expected to vanish in the limit of vanishing fermion mass.
As is well known, it turns out to be impossible for any regulator to enforce the conservation of  $j^{5\mu}$ at the one-loop quantum
level without violating gauge invariance, and a well-defined finite divergence $\pa_{\mu} j^{5\mu} = (N/16\pi^2)\, F \tilde F$ results
in a background gauge field $F_{\mu\nu}$ with $N$ massless fermions \cite{AdlJack}. This axial current divergence cannot be represented in terms
of the original gauge fields by any local effective action. If it could be, there would be the possibility of shifting the coefficient of this term
by a suitable constant and cancelling the anomaly. That this is impossible without violating gauge invariance implies that the axial anomaly is
a genuine quantum infrared (IR) effect, characterized by a non-local effective action in pure QED or QCD. The IR (or more properly
lightcone) implications of the anomaly are confirmed by the appearance of massless poles in anomalous amplitudes, signaling the existence 
of a low energy massless effective excitation and degree of freedom not present in the original classical Lagrangian \cite{GiaMot}.

Since the $U_A(1)$ symmetry is explicitly broken by the axial anomaly, any local EFT expansion assuming this symmetry will
necessarily miss the anomaly and its associated low energy massless excitation. Thus if one adopts a standard Wilsonian EFT approach
to low energy meson physics, expanding only in polynomials of local derivatives of the meson fields divided by a characteristic mass scale
of the order of $\Lambda_{_{QCD}}$ or the $\rho$-meson mass of $M_{\rho}\simeq 770\,$ MeV, no anomalous $U_A(1)$ contribution will be
present. This local EFT is necessarily incomplete, neither correctly predicting the $\eta'$ pseudoscalar meson mass nor the decay
constant of $\pi^0 \ra 2 \g$. Capturing these effects requires adding to the standard local EFT meson Lagrangian an additional term,
explicitly breaking the $U_A(1)$ symmetry, in a procedure that has come to be known as {\it anomaly matching} \cite{tHt,EFText}.

The necessity of modifying the standard EFT approach when quantum anomalies are present becomes clear when keeping track 
of the scaling dimensions of operators. All infrared relevant operators, scaling with dimensions $d \le D$ where $D=4$ is the
physical spacetime dimension must be included in the EFT. Local operators with $d > 4$ appear divided by $d-4$ powers of 
the mass scale, and being suppressed by $d\!-\!4$ powers of $k/M$ at low energies, are classified as irrelevant in the infrared.
Conversely, local operators with $d<4$, such as the Einstein-Hilbert action of classical GR (together with a possible
cosmological term) are infrared relevant. The case of logarithmic scaling corresponding to the marginal case $d=4$ 
requires special care. Since a logarithm cannot be represented as a local polynomial in $k$, it corresponds to a non-local
term in the effective action in position space. It should be included in the EFT since its low energy effects are not power law
suppressed by the UV mass scale $M_{\rho}$. Logarithmic scaling is exactly the behavior of the marginal operators in the 
quantum effective action generated by the anomaly \cite{MazMot}. The marginally relevant non-local contribution(s) in the 
effective action can have important non-trivial effects in the low energy EFT, including the appearance of new massless 
excitations in correlation functions, as the QED/QCD axial anomalies show \cite{GiaMot}.

Inclusion of non-local operators obtained from the quantum one-loop effects of massless degrees of freedom may still seem
counterintuitive from the standard Wilsonian EFT approach, viewed as integrating out only heavy degrees of freedom at mass-energy
scales greater than the UV scale $M$. Reference to the QCD axial anomaly will help to resolve any apparent paradox. In QCD
the mass scale of the light $u$ and $d$ quarks is of order $5$ to $10\,$ MeV. Since the meson EFT is supposed to be valid
up to the much larger UV cutoff scale for $M_{\rho}\simeq 770$ MeV, it follows that there is a large energy range
\be
\vspace{-1mm}
m_{u,d} \ll k \ll M_{\rho}
\vspace{-1mm}
\label{intE}
\ee
for which the light quarks are effectively massless. The decoupling theorem still holds for the very lowest energies $k \ll m_{u,d}$,
where all effects of the anomaly are suppressed by at least one factor of $(k/2 m_{u,d})^2$, but the existence of
mass scales much smaller than the UV cutoff scale $M_{\rho}$ means that anomalous quantum vacuum polarization effects
of the light quark loops survive in the meson EFT in the intermediate energy range (\ref{intE}), and the term in the EFT due
to the axial anomaly matching must be added to the low energy effective Lagrangian in this energy range. Thus even when
the fermion masses are non-zero, so that the underlying QFT has no exact chiral symmetry, the anomaly contributions
will be present in the range (\ref{intE}), where the light quark masses can be neglected.

In the case of gravity the UV cutoff at the Planck scale likewise does not preclude anomalous effects at much lower energies
because of the existence of additional light quantum fields which are either exactly massless, such as the photon, or
approximately so compared to the Planck scale. Thus over a large intermediate range of energies below the Planck
energy scale, semi-classical methods and an EFT approach to gravity including the effects of logarithmally scaling QFT anomalies
should apply.

In QFT in curved space the important quantity coupling to gravity is the matter stress-energy tensor $T^{\m\n}$. For massless fields 
conformally coupled to gravity, the stress tensor is classically traceless $T^\m_{\ \m} =0$. As in the case of the axial current, 
the classical conformal symmetry cannot be maintained at the one-loop quantum level without violating another invariance, 
in this case general coordinate invariance, expressed by the covariant conservation law $\na_\m T^{\m\n} = 0$. Enforcing the 
latter results in the trace being necessarily non-zero and given by
\vspace{-3mm}
\be
\lag T^\m_{\ \m} \rag = b\, C^2 + b' \left(E - \tfrac{2}{3}\sq R\right) + b'' \sq R + \sum_i \bet_i \,\cL_i
\vspace{-2mm}
\label{tranom}
\ee
in a general four dimensional curved spacetime, where
\vspace{-1mm}
\bes
\begin{align}
E &\equiv \,^*\hskip-.5mmR_{\a \bet\g \del}\,^*\hskip-.5mm R^{\a\bet\g\del} =
R_{\a\bet\g\del}R^{\a \bet \g \del}-4R_{\a\bet}R^{\a\bet} + R^2\\
C^2 & \equiv C_{\a \bet \g \del}C^{\a \bet \g \del} = R_{\a \bet \g \del}R^{\a \bet \g \del} -2 R_{\a\bet}R^{\a\bet}  + \tfrac{1}{3}R^2
\end{align}\label{EFdef}\ees
is the Euler-Gauss-Bonnet integrand and the square of the Weyl conformal tensor respectively. Here $R_{\a \bet \g \del}$
the Riemann curvature tensor, $^*\hskip-.5mm R_{\a \bet \g \del}$ its dual, and $R_{\a \bet}$ and $R$ is the Ricci tensor and scalar.
Additional dimension four invariants denoted by $\cL_i$ in (\ref{tranom}) may also appear in the general form of the trace anomaly,
if the massless conformal fields in question couple to additional long range fields. For example in the case of massless fermions
coupled to a gauge field, there are contributions from the scalar invariants $\cL_F= F_{\m\n}F^{\m\n}$ of electromagnetism or
of the strong or electroweak non-abelian gauge fields, with coefficients determined by the $\beta$ function of the corresponding
gauge coupling. Note that these gauge field terms are certainly {\it not} Planck suppressed.

All the anomaly coefficients $b$, $b', b''$ and $\bet_i$ are dimensionless parameters multiplied by $\hbar$, and
\bes
\bea\vspace{-1mm}
b\,&=&\, \frac{\hbar}{120 (4 \pi)^2}\ \big(N_s + 6 N_f + 12 N_v\big)\\
b'&=& -\frac{\hbar}{360 (4 \pi)^2}\, \big(N_s + 11 N_f + 62 N_v\big)
\vspace{-1mm}
\label{bprime}
\eea
\label{bbprime}\ees
are known for any number of free conformal scalars ($N_s$), four-component Dirac fermions ($N_f$) or vectors ($N_v$)
respectively  \cite{BirDav,Duff}.  The $b''$ coefficient in (\ref{tranom}) is ultraviolet (UV) regularization dependent and can
be changed at will or set to zero by the addition of a local $R^2$ term in the effective action, and hence is not a true anomaly.
In contrast, the $b$ and $b'$ terms in (\ref{tranom}) are independent of the UV regulator, cannot be removed by any local
counterterm, and depend on the number of massless fields of each spin, which is a property of the low energy effective
description of matter coupled to gravity. Note that $b>0$ but $b'<0$ for all non-gravitational free fields of spin less than
or equal to one.

Since the terms in the quantum effective action of the trace anomaly, {\it cf.} Eq. (\ref{Sanom}) below, scale logarithmically
with distance, they may be regarded as marginally relevant terms in the low energy Wilsonian effective action for gravity,
with the $b, b'$ coefficients treated as arbitrary parameters to be fixed by experiment, and to be added to the usual
$R$ and $\La$ terms of the Einstein-Hilbert action in a local EFT expansion in derivatives. From this latter point of view
the effective action corresponding to the trace anomaly is a non-trivial extension, or modification of Einstein's classical theory
at energies far below the Planck scale, and is hence relevant for macroscopic physics. Unlike most other possible modifications
of GR, the extension by a term associated with the conformal anomaly is {\it required} by known one-loop quantum vacuum 
polarization effects of light fields in general curved spaces, and being generally covariant is consistent with the Equivalence Principle,
at least in its weak sense.

Note also that inclusion of the effects of the trace anomaly does not rely in any way upon the classical
Einstein-Hilbert action or the value of the gravitational constant $G$ itself. The Einstein eqs. are not used in deriving
(\ref{tranom}) for an arbitrary classical background metric. As in the case of the QCD axial anomaly, the underlying QFT
need not be exactly conformally invariant, or the axial and/or conformal invariance may be spontaneously broken.
Notwithstanding any of the fields contributing to $b$ and $b'$ in (\ref{bbprime}) having a finite mass $m\ll M_{Pl}$, their
anomalous vacuum contributions to the effective action must still be taken into account for intermediate energies
$m \ll k \ll M_{Pl}$, in the low energy EFT of gravity far below the Planck scale \cite{MazMot,MotVau,Zak}.

It is this semi-classical EFT, consisting of the usual Einstein-Hilbert action of classical General Relativity together with the quantum
effective action generated by the conformal trace anomaly, Eqs. (\ref{Sanom}) and (\ref{Seff}) below, which lead to an additional
massless dynamical scalar in the conformal sector, and scalar gravitational waves that propagate over macroscopic distances,
that is the main subject of this paper.

The paper is organized as follows. In the next section the effective action of the conformal trace anomaly  
and the Wess-Zumino consistency condition it satisfies are briefly reviewed in both its non-local and equivalent local form, 
the latter by the introduction of a scalar field $\vf$, to be called the {\it conformalon} field. The stress-energy tensor of the scalar 
conformalon field is given in Sec.\,\ref{Sec:SET}. The resulting EFT containing both metric and scalar conformalon $\vf$ field 
is linearized around flat spacetime in Sec.\,\ref{Sec:LinFlat}, and gauge invariant metric variables introduced in both a covariant 
and a $3 + 1$ spacetime splitting, demonstrating the coupling of $\vf$ to the conformal part of the metric at linear order. 
In Sec.\,\ref{Sec:Green} the scalar Green's function for $\vf$ is used to express the solutions in terms of the strength of localized 
sources. The energy and power radiated in scalar gravitational waves is computed to quadratic order in the fluctuations and
given in terms of the source strength in Sec.\,\ref{Sec:EnerFlux}. Sec.\,\ref{Sec:Astro} contains some preliminary estimates 
of the strength of scalar gravitational waves from astrophysical sources, including the possibility of detection by present or 
planned gravitational wave antennas. Sec.\,\ref{Sec:Conc} contains a Summary of the results and Outlook for future work
on the EFT modification of General Relativity by the anomaly effective action.

\section{The Conformal Anomaly and Low Energy EFT of Gravity}
\label{Sec:EFT}

The derivation of the effective action encapsulating the trace anomaly (\ref{tranom}) is straightforward.
Introducing the local conformal parameterization of the metric
\be
g_{\m\n}(x) = e^{2\s (x)} \bar g_{\m\n}(x) \,,\qquad \sqrt{-g_x} \equiv [-{\rm det}\,g_{\m\n}(x)]^{\frac{1}{2}} = e^{4\s (x)}\,\sqrt{-\bar g_x}
\label{confdef}
\vspace{-2mm}
\ee
in terms of an arbitrary fixed fiducial metric $\bar g_{\m\n}$, the conformal dependences of the terms
in (\ref{tranom}) are
\bes
\bea
\sqrt{-g}\,C^2 &=& \sqrt{-\overline g}\,\overline C^2\,\label{Csig}\\
\sqrt{-g}\,\cL_i &=& \sqrt{-\overline g}\,\overline\cL_i \label{Lsig}\\
\sqrt{-g}\,\left(E - \tfrac{2}{3}\sq R\right) &=& \sqrt{-\overline g}\,
\left(\overline E - \tfrac{2}{3}\sqb\overline R\right) + 4\,\sqrt{-\overline g}\, \bar\D_4\,\s  \label{Esig}
\eea
\label{CEsig}
\ees
\vskip-1cm
\noindent where the coordinate label $_x$ subscript on $\sqrt{-g_x}= \sqrt{-g}$ is generally suppressed
when it causes no confusion to do so. All quantities with an overbar are evaluated in the fixed fiducial
metric $\bar g_{\m\n}$. The fourth order differential operator \cite{Rie,PanBran,AntMot,MazMot}
\be
\D_4 \equiv \na_\m \left(\na^\m\na^\n +2R^{\m\n} - \tfrac{2}{3} R g^{\m\n} \right)\na_\n
=\sq^2 + 2 R^{\m\n}\na_\m\na_\n - \tfrac{2}{3} R \sq + \tfrac{1}{3} (\na^\m R)\na_\m
\label{Deldef}
\ee
is the unique fourth order scalar kinetic operator that is conformally covariant
\vspace{-2mm}
\be
\sqrt{-g}\, \D_4 = \sqrt{-\bar g}\, \bar \D_4
\label{invfour}
\ee
for arbitrary $\s(x)$.

From (\ref{CEsig}) it is clear that of the various terms in the general form of the trace anomaly (\ref{tranom}), the particular linear
combination  of the topological density $E$ and $-\frac{2}{3}\sq R\,$ in (\ref{Esig}) plays a special role. It is this linear
combination and this combination only whose conformal variation is linear in $\s$ and defines the differential operator
$\D_4$. This is a general feature in any even $d=2n$ dimensional spacetime, as there exists a particular set of local invariants
which when added to the Euler-Gauss-Bonnet topological density in any even dimension defines a conformally covariant $(2n)^{th}$
order scalar differential operator operating linearly on $\s$.  All other terms in the general form the conformal anomaly (\ref{tranom})
are either conformally invariant as (\ref{Csig})-(\ref{Lsig}) or cohomologically trivial, as is $\sq R$ by itself, in the sense that they can
be removed by appropriate local counterterms in the effective action.

If (\ref{tranom}) is multiplied by $\sqrt{-g}$ the result is the conformal $\s$ variation of a Wess-Zumino effective action
\vspace{-7mm}
\be
\frac{\delta \G_{_{\!W\!Z}}} {\del \s}\bigg\vert_{\bar g} = \sqrt{-g} \, \lag T^\m_{\ \m}(x) \rag \equiv \cA
\label{cAdef}
\ee
which gives rise to the anomalous trace. Because of (\ref{confdef}) the linear $\s$ dependence in (\ref{CEsig}) is a {\it logarithmic}
scaling with metric distances, which makes $\G_{_{\!W\!Z}}$ a marginally relevant effective action. This Wess-Zumino effective
action is easily found by inspection of the linear relations (\ref{CEsig}), {\it viz.} \cite{Rie}
\bea
\hspace{-1mm}
&&\G_{_{\!W\!Z}}[\bar g;\s] = 2 b'\! \int\,d^4x\,\sqrt{-\bar g_x}\ \s\,\overline\Delta_4\,\s + \int\,d^4x\ \overline{\!\cA} \, \s\nn
&&\quad = b' \int\,d^4x\,\sqrt{-\bar g_x}\,\Big[2\,\s\,\bar\Delta_4\,\s + \left(\bar E - \tfrac{2}{3} \sqb \bar R\right)\s \Big]_x
+ \int\,d^4x\,\sqrt{-\bar g_x}\, \Big[b\,\bar C^2\,\s + \sum_i \bet_ i\,{\bar\cL}_i\,\s\Big]_x
\label{WZfour}
\eea
up to conformally invariant terms terms independent of $\s$, which being insensitive to rescalings do not generate
infrared relevant terms. The action $\G_{_{\!W\!Z}}$ satisfies the Wess-Zumino consistency condition \cite{WZ,AMMNP}
\vspace{-7mm}
\be
\G_{_{\!W\!Z}}[\bar g;\s]  = \G_{_{\!W\!Z}}[e^{2\w}\bar g;\s - \w]  + \G_{_{\!W\!Z}}[\bar g;\w] \,.
\label{WZcons}
\ee
By its construction the WZ action $\G_{_{\!W\!Z}}$ depends separately on $\s$ and the arbitrary background
metric $\bar g_{\m\n}$ and is therefore not fully generally coordinate invariant. However, owing to (\ref{WZcons})
$\G_{_{\!W\!Z}}$ is a cohomological representation of the group of local Weyl transformations of the metric, and
consequently can be expressed as a difference of fully covariant but non-local effective actions evaluated at the metric
$g_{\m\n}(x)$ and $\bar g_{\m\n}(x)$ \cite{MazMot}. This non-local difference form
\be
\G_{_{\!W\!Z}}[\bar g;\s] = \cS_{anom}^{^{NL}}[g=e^{2\s}\bar g] - \cS_{anom}^{^{NL}}[\bar g],
\label{Weylshift}
\ee
is obtained by solving (\ref{Esig}) formally for $\s$, thereby inverting the differential operator $\D_4$, substituting the
result into (\ref{WZfour}) and using (\ref{invfour}). The resulting covariant non-local effective action is given by
\be
\cS_{anom}^{^{NL}}[g] =\sdfrac {1}{4}\!\int \!d^4x\sqrt{-g_x}\, \big(E - \tfrac{2}{3}\sq R\big)_{\!x} \!
\int\! d^4x'\sqrt{-g_{x'}}\,D_4 (x,x')\bigg[\sdfrac{b'}{2}\, \big(E - \tfrac{2}{3}\sq R\big) +  b\,C^2 + \sum_i \bet_i\,\cL_i\bigg]_{x'}
\label{Snonl}
\ee
where $D_4(x,x')$ denotes the Green's function inverse of the fourth order differential operator
$\D_4$ defined by (\ref{Deldef}), in the sense that
\vspace{-5mm}
\be
\int d^4 x' \, \sqrt{-g_{x'}}\ \Delta_4\,D_4(x,x')\, \psi(x') = \psi(x)
\ee
for any scalar function $\psi(x)$. The particular Green's function depending on appropriate boundary conditions must be selected 
by the physical application. In Sec. \ref{Sec:Green} we solve for the retarded Green's function $D_{4, Ret} \equiv D_R(x,x')$
explicitly with classical retarded boundary conditions in flat space. Similar to the retarded Green's function for the usual second
order wave operator $\sq$ in two dimensions, $D_R(x,x')$ is a constant for all points $x'$ within the entire past light cone of $x$,
and the corresponding Feynman propagator grows logarithmically at large distances. This feature allows the effects of the
non-local quantum correlations induced by the conformal trace anomaly (\ref{tranom}), contained in the anomaly
effective action to be cumulative and non-negligible at macroscopic distance scales. Adding a $\sq R$ term to
(\ref{tranom}) with an arbitrary $b'' \neq 0$ would add a local $R^2$ term to the effective action, which affects
the UV behavior but which only produces Planck suppressed effects at low energies, without altering the non-local IR
contribution (\ref{Snonl}) of the true anomaly.

It bears emphasizing that the anomaly effective action (\ref{Sanom}) is completely distinct from all terms local in the curvature 
and its derivatives in EFT, as well as any other possible non-local but conformally invariant terms that might be generated by 
integrating out quantum matter fields, in both its logarithmic scaling with distance, and the special role of the $\D_4$ operator 
in producing a particular {\it kinetic energy term} for $\s$. This kinetic energy term in the Wess-Zumino action (\ref{WZfour})
shows that the conformal part of the metric becomes {\it dynamical} in the low energy EFT of gravity, whereas it is constrained
in classical General Relativity. The covariant non-local form (\ref{Snonl}) of the anomaly effective action shows that the
dynamical degree of freedom in the conformal sector is not associated with the Planck scale, quite unlike possible additional 
degrees of freedom introduced by local higher derivative curvature terms. The effects of such higher local higher derivative 
terms have been considered in \cite{DonBur}, and are suppressed by $(L_{pl}/ r)^2$ where $r$ is a typical macroscopic scale. 
This need not be the case for (\ref{Snonl}), the effects of which are non-local in the invariants including $\cL_i$, and which are 
not uniformly suppressed by the Planck scale. The effective semi-classical action (\ref{Snonl}) in the gravitational sector, in which 
$b, b'$ proportional to $\hbar$ appear as parameters, may be regarded as an efficient bookkeeping device to take into account 
the anomalous vacuum polarization effects of the known light fields in the Standard Model in the gravitational sector, without 
having to calculate loops containing these light fields in every individual process.

In order to make the scalar excitation associated with the anomaly and its macroscopic consequences manifest in a coordinate 
independent way, it is convenient to recast the generally covariant non-local effective action (\ref{Snonl}) in local form by the 
introduction of (at least one) scalar field. Because it is asymmetrical in the invariants $E$ and $(C^2, \cL_i)$, two additional 
scalar fields would be necessary to render the minimal non-local action (\ref{Snonl}) into a local form \cite{MotVau}. On the 
other hand, if one adds to the effective action (\ref{Snonl}) the Weyl invariant terms necessary to symmetrize in
$x$ and $x'$, and complete the square, namely if one adds to (\ref{Snonl}) the terms
\be
\sdfrac {1}{8b'}\!\int \!d^4x\sqrt{-g_x}\, \Big[ bC^2 + \sum_i \beta_i \cL_i \Big]_{x} \int\! d^4x'\sqrt{-g_{x'}}\,D_4(x,x') 
\Big[ bC^2 + \sum_i \beta_i \cL_i \Big]_{x'}
\label{SaddW}
\ee
then one obtains the symmetric non-local form of the anomaly effective action
\be
\cS_{\rm anom}^{^{NL}}[g] \rightarrow  \sdfrac {1}{8b'}\int d^4x \int d^4x'\, \cA(x)\, D_4(x,x')\, \cA(x')
\label{Snonlsq}
\ee
where the total trace anomaly density $\cA$ in (\ref{cAdef}) is given by (\ref{tranom}) with $b'' =0$. Since the
anomaly effective action $\cS_{\rm anom}^{^{NL}}$ is determined only up to Weyl invariant terms in any case,
adding Weyl invariant terms such as (\ref{SaddW}) to it does not affect the trace anomaly or (\ref{Weylshift}). 
However adding a term such as (\ref{SaddW}) does change the {\it tracefree} parts of the stress tensor
derivable from the action in (\ref{Snonl}) {\it vs.} (\ref{Snonlsq}), and whether the effective action contains
non-minimal Weyl invariant terms such as (\ref{SaddW}) or not can only be determined by explicit calculation 
in particular QFTs. The results of this paper are insensitive to whether the anomaly action is symmetrized by
adding (\ref{SaddW}) or not.

Assuming for simplicity the symmetrized form (\ref{Snonlsq}), one can now introduce a single scalar field $\vf$, requiring 
it to satisfy the linear equation of motion
\be
\D_4\, \vf = \sdfrac{E}{2}- \sdfrac{\!\sq R\!}{3} + \sdfrac{1}{2b'} \Big(b\,C^2 +  \sum_i \bet_i\cL_i\Big) = \sdfrac{1}{2b'\sqrt{-g}}\ \cA
\label{phieom}
\ee
and rewrite the non-local action (\ref{Snonlsq}) in its equivalent local form
\bea
&&S_{anom}[g;\vf] \equiv -\sdfrac{b'}{2\,} \int d^4x\sqrt{-g}\, \bigg[ (\sq \vf)^2 - 2 \big(R^{\m\n} - \tfrac{1}{3} R g^{\m\n}\big)
(\na_\m\vf)(\na_\n \vf)\bigg]\nn
&& \hspace{1.5cm} +\, \sdfrac{1}{2}\,\int d^4x\sqrt{-g}\  \bigg[
b'\left(E - \sdfrac{2}{3}\sq R\right) + b\,C^2 + \sum_i \bet_i \cL_i\bigg]\,\vf
\label{Sanom}
\eea
by using (\ref{Deldef}) and integrating by parts. The advantage of this local form of $S_{anon}[g;\vf]$ is that free variation with respect 
to $\vf$ yields back its equation of motion (\ref{phieom}), and is otherwise entirely equivalent to the non-local form (\ref{Snonlsq})
which is reproduced (up to surface terms) if (\ref{phieom}) is solved for $\vf$ by inverting $\D_4$ and the result is substituted into 
(\ref{Sanom}). The freedom to change the boundary conditions on the Green's function inverse $D_4(x,x')$ of the wave operator 
$\D_4$ by adding homogeneous solutions of this wave operator to $D_4(x,x')$ in the non-local effective action is equivalent to the 
freedom in the initial data to solve (\ref{phieom}) for $\vf$ in the local form (\ref{Sanom}). The kinetic terms of the local scalar $\vf$ 
in (\ref{Sanom}) are mapped to the massless scalar propagator anomaly poles in the non-local form (\ref{Snonlsq}), which have been 
verified explicitly in three-point correlation functions as a necessary consequence of anomalous Conformal Ward Identities of otherwise 
conformal field theories \cite{GiaMot,CorMagMot}. 

Note that unlike $\G_{_{\!W\!Z}}[\bar g;\s]$, $S_{anom}[g;\vf]$ depends upon the full physical metric $g_{\m\n}$ and is therefore
fully general coordinate invariant. It is clear that $\vf$ is nevertheless very closely related to the conformal factor $\s$ of the metric itself 
in (\ref{confdef}). Indeed
\vspace{-2mm}
\be
S_{anom}\big[g;\vf\big] = - \G_{_{\!W\!Z}} \left[g; - \sdfrac{\vf}{2}\right]
\label{SanomGam}
\ee
so that (\ref{WZcons}) implies
\vspace{-4mm}
\be
S_{anom}[e^{-2 \s} g; \vf] = S_{anom}[g; \vf + 2 \s] - S_{anom}[g; 2 \s]
\vspace{-1mm}
\label{SanomWZ}
\ee
which shows that conformal transformations of the metric are related to linear shifts in the spacetime scalar $\vf$.
For this reason and because the identity (\ref{SanomWZ}) exposes its origin and fundamental relationship to variations
of the conformal frame of the metric, $\vf$ may be termed the scalar {\it conformalon}\,  field, which 
serves to distinguish it from dilatons and dilaton-like fields that arise in other contexts. Because of the fourth
order kinetic term the scalar conformalon $\vf$ has canonical mass dimension zero, which also distinguishes
it from other dimension one dilaton-like fields.

The method of obtaining (\ref{Sanom}) outlined relies upon integrating the trace anomaly eq.~(\ref{tranom}) with the values
of the coefficients $b, b', \bet_i$ dependent upon the underlying QFT content. In that case the scalar conformalon $\vf$
does not introduce any genuinely `new' degrees of freedom, but rather re-expresses in a convenient form
certain two-particle correlations present in the underlying QFT vacuum \cite{GiaMot,CorMagMot,BlaCabMot}. For example,
the Casimir effect and vacuum polarization effects in fixed geometries such as the Schwarzschild and de Sitter backgrounds may be
computed for a large variety of states for fields of arbitrary spin by use of $S_{anom}$ and its corresponding stress tensor
(\ref{Eab}) \cite{MotVau}. On the other hand, since the {\it form} of the trace anomaly (\ref{tranom}) is prescribed by locality
of QFT and general coordinate invariance, the effective action (\ref{Snonl}) associated with it may be regarded as a necessary
part of the general effective action of low energy gravity, based on dimensional scaling and invariance principles alone.
From that perspective the existence of (at least one) scalar field $\vf$ coupling to the spacetime metric and satisfying 
(\ref{SanomWZ}) is independent of any specific matter field content, and the coefficients $b, b', \bet_i$ and of (\ref{SaddW})
may be treated as free parameters of low energy gravity, to be determined by experiment.

When (\ref{Sanom}) is added to the usual Einstein-Hilbert term of classical General Relativity
\be
S_{EH}[g] = \frac{1}{16\pi G} \int\, d^4x\,\sqrt{-g}\, \big(R-2\La\big) 
\label{Scl}
\ee
we obtain the semi-classical effective action
\vspace{-3mm}
\be
S_{eff}[g;\varphi] = S_{EH}[g] + S_{anom}[g;\vf]
\label{Seff}
\vspace{-1mm}
\ee
which defines a scalar-tensor low energy effective theory of gravity. Since $\vf (x)$ is a spacetime scalar field, the trace
anomaly action (\ref{Sanom}) is invariant under general coordinate transformations, just as the classical action (\ref{Scl}) is.
As a metric theory that leaves unaltered the direct minimal gravitational couplings of matter through covariant derivatives,
the effective action (\ref{Seff}) is thus consistent with the Einstein Equivalence Principle, by admitting local freely falling
Lorentz frames at every point where the non-gravitational laws of physics take on their familiar Lorentz invariant form.

Because the scalar conformalon couples to gauge fields through the $\bet_i\cL_i$ terms in the trace anomaly and
influences the metric and gravitational interactions through its stress tensor, {\it cf.}~(\ref{Ttens}) below, certain more
subtle violations of the Strong Equivalence Principle (SEP) are possible in the EFT defined by (\ref{Seff}). Since the 
scalar $\vf$ is the minimal one required by the general form of the quantum trace anomaly (\ref{tranom}) and general 
covariance, the addition of $S_{anom}$ and any violations of the SEP it generates are the minimal ones required by 
general principles of QFT. In this connection one should note that both the form and the fundamental origin of (\ref{Sanom}) 
in the correlations of massless quantum fluctuations differ markedly from other scalar-tensor theories, such as 
Fierz-Jordan-Brans-Dicke (FJBD) theory \cite{BrDicke}. In particular, the scalar conformalon $\vf$ does {\it not} couple 
to the full trace of the classical stress tensor of massive matter, as the FJBD scalar does, but only linearly to those dimension 
four conformal invariants with no dimensionful mass parameters, such as (\ref{Lsig}), which are dictated by the quantum 
trace anomaly (\ref{tranom}) and Wess-Zumino consistency (\ref{SanomWZ}). Additional self-couplings of $\vf$ would 
generally violate WZ consistency. Since the possible sources for $\vf$ in (\ref{phieom}) are negligibly small in our local 
neighborhood, this also allows (\ref{Seff}) to easily pass the stringent solar system tests which constrain the coupling(s) 
in FJBD theory, as well as the laboratory constraints on other modified gravity theories \cite{Will}.

The presence of four derivatives in $\D_4$ naturally raises questions about unitarity and ghosts 
in the quantum EFT of (\ref{Sanom}). In the context of the pure anomaly or `free' theory defined by (\ref{WZfour}), quantized on 
${\mathbb R} \times {\mathbb S}^3$, the negative norm ghost states are eliminated from the physical spectrum by the 
diffeomorphism constraints \cite{AMMStates}. The proper framework for the resolution of the question of ghosts 
in the EFT of low energy gravity (\ref{Seff}) requires making full use of the (four) constraints of diffeomorphism invariance. 
Because of (\ref{SanomWZ}), $\vf$ mixes with the conformal part of the metric, and the scalar conformalon $\vf$ is always partly 
constrained, as the conformal part of the metric is fully constrained in classical GR, {\it cf.} Sec. \ref{Sec:LinFlat}. Since 
(\ref{SanomWZ}) forbids adding arbitrary additional non-linear interactions in the conformalon sector, it is only the non-linearity 
of classical GR itself that prevents the constraints from being solved in closed form, but these non-linear couplings are suppressed 
in flat space by the weakness of the gravitational coupling $G$, up to Planck energy scales where the EFT framework is expected
to break down. Since in this paper only {\it linearized} perturbations about flat space are considered, which may be treated by 
{\it classical} methods, with $\hbar$ entering only through the $b, b', \beta_i$ coefficients of the anomaly, but with the scalar conformalon
$\vf$ taken to be a c-number field, the results do not rely upon quantization of $\vf$, and are independent of questions or concerns 
about the quantum EFT, unitarity and ghosts, which will be addressed in a separate publication. 

In \cite{DSAnom} the full effective action (\ref{Seff}) with a cosmological term was considered perturbatively in linear response
around de Sitter space. In that case the higher derivative mode of $\D_4$ decouples if one restricts attention to excitations far below
the Planck scale, for which the EFT description is valid, but a physical scalar mode of $\vf$ with {\it second order kinetic term}
not present in the classical Einstein theory survives. In this paper this analysis is carried out for the somewhat simpler case of 
linear perturbations around flat space, with the result that again a single {\it second order} physical scalar excitation survives 
and couples to the metric at low energies, which with {\it positive} energy is stable, and is responsible for the new phenomenon 
of scalar gravitational waves, treated classically in the EFT of (\ref{Seff}).
\vspace{-3mm}

\section{Stress-Energy of the Scalar Conformalon}
\label{Sec:SET}
\vspace{-2mm}

Variation of the last form of the anomaly action (\ref{Sanom}) with respect to the metric yields the stress-energy tensor
of the scalar conformalon field
\be
T_{\m\n}[\vf] \equiv -\frac{2\!\!\!}{\!\!\sqrt{-g}}\, \frac {\del }{\del g^{\m\n}} \,S_{anom}[g;\vf] =
b'\, E_{\m\n} + b\, C_{\m\n} + \sum_i \bet_i \,T^{(i)}_{\m\n}[\vf]
\label{Ttens}
\ee
where
\vspace{-4mm}
\bea
&&E_{\m\n} \equiv  - 2\,(\na_{(\m}\vf) (\na_{\n)} \sq \vf)  + 2\na^\a \big[(\na_\a \vf)(\na_\m\na_\n\vf)\big]
- \tfrac{2}{3}\, \na_\m\na_\n\big[(\na_\a \vf)(\na^\a\vf)\big]\nn
&&\hspace{1.2cm}+\tfrac{2}{3}\,R_{\m\n}\, (\na_\a \vf)(\na^\a \vf) - 4\, R^\a_{\ (\m}\left[(\na_{\n)} \vf) (\na_\a \vf)\right]
 + \tfrac{2}{3}\,R \,(\na_{(\m} \vf) (\na_{\n)} \vf) \nn
&&
\hspace{.5cm} + \tfrac{1}{6}\, g_{\m\n}\, \left\{-3\, (\sq\vf)^2 + \sq \big[(\na_\a\vf)(\na^\a\vf)\big]
+ 2\, \big( 3R^{\a\bet} - R g^{\a\bet} \big) (\na_\a \vf)(\na_\bet \vf)\right\}\nn
&&\hspace{-7mm}  - \tfrac{2}{3}\, \na_\m\na_\n \sq \vf  - 4\, C_{\m\ \n}^{\ \a\ \bet}\, \na_\a \na_\bet \vf
- 4\, R_{(\m}^\a \na_{\n)} \na_\a\vf  + \tfrac{8}{3}\, R_{\m\n}\, \sq \vf   +\tfrac {4}{3}\, R\, \na_\m\na_\n\vf
- \tfrac{2}{3} \left(\na_{(\m}R\right) \na_{\n)}\vf\nn
&& \hspace{1.8cm} + \tfrac{1}{3}\, g_{\m\n}\, \left[ 2\, \sq^2 \vf + 6\,R^{\a\bet} \,\na_\a\na_\bet\vf
- 4\, R\, \sq \vf  + (\na^\a R)\na_\a\vf\right]
\label{Eab}
\eea
is the metric variation of the terms proportional to $b'$, both quadratic and linear in $\vf$ in (\ref{Sanom}), and
\bes
\bea
&&C_{\m\n} \equiv  -\frac{2}{\sqrt{-g}\ } \frac {\del }{\del g^{\m\n}} \left\{\frac{1}{2} \int d^4x\sqrt{-g}\,C^2\,\vf\right\}
=-4\,\na_\a\na_\bet\big( C_{(\m\ \n)}^{\ \ \a\ \bet} \,\vf \big)  -2\, C_{\mu\ \nu}^{\ \a\ \bet}\, R_{\a\bet}\, \vf\\
&& \hspace{2.5cm}T^{(i)}_{\m\n}[\vf] \equiv  -\frac{2}{\sqrt{-g}\ }\, \frac{\del}{\del g^{\m\n}}\left\{\frac{1}{2} \int d^4x\sqrt{-g}\,  \cL_i\, \vf\right\}
\eea
\label{CFterms}\ees
are the metric variations of the last two $b$ and $\bet_i$ terms in (\ref{Sanom}), linear in $\vf$. Thus for example,
\be
T^{(F)}_{\m\n}[\vf] = \big(\!-2F_\m^{\ \a}\,F_{\a\n} + \, \tfrac{1}{2}\,g_{\m\n} F^{\a\bet}F_{\a\bet}\big)\,\vf
\label{TMax}
\ee
for  $\cL= F_{\m\n}F^{\m\n}$, which is proportional to $\vf$ and the Maxwell stress tensor.

Since the effective action is a coordinate invariant scalar, the stress-energy tensor (\ref{Ttens}) is covariantly conserved
\vspace{-5mm}
\be
\na_\m T_{\ \,\n}^{\m}[\vf] = 0
\ee
for any  $b, b'$ upon making use of the $\vf$ eq. of motion (\ref{phieom}), and for any $\bet_i$, if supplemented by the eqs.
of motion for any additional fields coupled through $\cL_i$. It is easily verified that the terms quadratic in $\vf$ in (\ref{Eab}),
as well the terms (\ref{CFterms})-(\ref{TMax}) are conformally invariant and traceless, so that the total trace of the anomaly
stress tensor is obtained from the terms in (\ref{Eab}) linear in $\vf$, yielding
\be
g^{\m\n} T_{\m\n}[\vf] = 2b'\,\D_4 \vf = b\, C^2 + b' \left(E - \tfrac{2}{3}\sq R\right) + b'' \sq R + \sum_i \bet_i\,\cL_i
\ee
which reproduces (\ref{tranom}), by again using the classical eq. of motion (\ref{phieom}). This relation may also be derived
as a direct consequence of Wess-Zumino consistency (\ref{SanomWZ}).

The stress tensor (\ref{Ttens}) appears as an additional source to the Einstein field equations obtained by varying
the full gravitational effective action (\ref{Seff}) with respect to the metric, which thus become
\be
R_{\m\n} - \tfrac{1}{2} g_{\m\n} R + \La g_{\m\n} = 8 \pi G\, \left( T^{(cl)}_{\m\n} + T_{\m\n} [\vf]\right)
\label{Ein}
\ee
in the EFT. Here $T^{(cl)}_{\m\n}$ denotes the classical stress-energy tensor of all other matter and radiation fields. Thus the semi-classical
EFT consists of classical GR supplemented by a well-defined additional conserved stress-energy tensor which encapsulates the
vacuum fluctuation and polarization effects associated with the conformal trace anomaly of light or massless quantum fields.
The form and value of (\ref{Ttens}) given by (\ref{Eab})-(\ref{TMax}) depend in its tracefree parts quadratic in
$\vf$ upon the completion of the square by addition of the Weyl invariant terms (\ref{SaddW}). If these Weyl invariant
terms are absent from the full effective action, the slightly more complicated two-field effective action and stress tensor of the 
anomaly given in \cite{MotVau} should be used in place of the tracefree quadratic terms in (\ref{Ttens}) and (\ref{Eab}).

\section{Linearization Around Flat Space}
\label{Sec:LinFlat}
\subsection{Covariant Tensor Decomposition}

It is clear that with $\La =0$ the modified field eqs.~(\ref{Ein}) with (\ref{phieom}) are satisfied by the vacuum solution
$\vf =0$ in flat space $g_{\m\n} = \eta_{\m\n}$ with no other sources. An important physical consequence of adding
the anomaly action to classical GR can be seen at linearized order around this flat spacetime vacuum solution.
The effective action (\ref{Seff}) expanded to second order in the field fluctuations
\be
h_{\m\n} \equiv g_{\m\n} - \eta_{\m\n} \,, \qquad h \equiv \eta^{\m\n}h_{\m\n}
\label{hdef}
\ee
around the vacuum solution of flat spacetime with $\eta_{\m\n} = {\rm diag} (-+++)$ and $\vf =0$ is
\bea
&&S^{(2)}_{eff}\big\vert_{g=\eta} = \sdfrac{1}{32 \pi G}  \int\, d^4x\, \left\{\tfrac{1}{2} h^{\m\n} \sq h_{\m\n} + h\, \na^\m\na^\n h_{\m\n}
+ (\na^\n h_{\m\n})( \na^\a h_\a^{\  \m}) -\tfrac{1}{2} h \sq h\right\}\nn
&&\hspace{1.5cm}  -\sdfrac{b'}{2\,} \int\, d^4x\ (\sq \vf)^2 + \sdfrac{b'}{3\,} \int\, d^4x\ (\sq \vf)\,\big\{\sq h - \na^\m\na^\n h_{\m\n}\big\}
\label{Seff2}
\eea
where surface terms have been discarded. The last term in (\ref{Seff2}) shows that there is a mixing
between the trace of the metric perturbation $h_{\m\n}$ and the scalar conformalon $\vf$. Note also that this mixing
of $\vf$ to the metric perturbation is only through $\sq \vf$. Thus any scalar perturbations with
$\sq \vf = 0$ are completely decoupled from the metric perturbations at linear order.

One can analyze the coupling between the anomaly scalar with $\sq \vf \neq 0$ and the linearized metric perturbation in either a
covariant or a canonical framework. In a covariant treatment space and time indices are treated on an equal footing,
so that Lorentz invariance is manifest, whereas in a canonical treatment spacetime is split into space + time, and true
propagating modes containing kinetic terms may be more clearly separated from constrained modes.

Considering first the spacetime covariant decomposition of the metric perturbation
\be
h_{\m\n} = u_{\m\n}^{\perp} + \pa_\m v_\n^{\perp} + \pa_\n v_\m^{\perp} + \left(\pa_\m\pa_\n - \tfrac{1}{4} \eta_{\m\n} \sq\right) w
+ \tfrac{1}{4} \eta_{\m\n} h
\label{metdecomp}
\ee
where $u_{\m\n}^{\perp}$ and $v_\m^{\perp}$ are transverse and $u_{\m\n}^{\perp}$ is traceless in the
four-dimensional covariant sense, {\it i.e.}
\bes
\bea
&&\pa^\m u_{\m\n}^{\perp}  = 0 = \eta^{\m\n} u_{\m\n}^{\perp} \\
&&\pa^\m v_\m^{\perp} = 0
\eea
\ees
then under the infinitesimal coordinate transformation $x^\m \rightarrow x^\m + \xi^\m(x)$
\be
\begin{aligned}[l]
&u_{\m\n}^{\perp} \rightarrow u_{\m\n}^{\perp}\\
&v_\m^{\perp} \rightarrow v_\m^{\perp} + \xi_\m^{\perp}
\end{aligned}
\hspace{4cm}
\begin{aligned}[r]
&w \rightarrow w + 2 \z\\
&h \rightarrow h + 2 \sq \z
\end{aligned}
\label{gaugetranscov}\ee
where $\xi_\m = \xi_\m^{\perp} + \pa_\m \z$ is also decomposed into transverse and longitudinal components
in the same covariant four-dimensional sense. Thus, of the ten independent metric components of the
linearized metric perturbation $h_{\m\n}$ only the subset of
\bes
\bea
u_{\m\n}^{\perp} &\rightarrow &u_{\m\n}^{\perp}\\
h - \sq w &\rightarrow & h -\sq w
\vspace{-3mm}
\eea
\label{gaugeinvcov}\ees
consisting of $5$ components of a spin-$2$ tensor field and $1$ component of spin-$0$ scalar field are gauge invariant
under coordinate transformations (up to possible zero modes). Accordingly, since the effective action (\ref{Seff}) is generally
coordinate invariant, the equations of motion (\ref{Ein}) and the second variation (\ref{Seff2}) can depend upon only these $6$
gauge invariant metric components. Indeed the Riemann and Einstein tensors, and the Ricci scalar linearized around flat space
are respectively
\bes
\bea
\del R_{\m\n\a\bet} &=&\del R_{\m\n\a\bet}^{(T)} + \del R_{\m\n\a\bet}^{(S)} = \tfrac{1}{2} \left\{ \pa_\a\pa_\n u_{\m\bet}^{\perp}
+ \pa_\bet\pa_\m u_{\n\a}^{\perp}- \pa_\bet\pa_\n u_{\m\a}^{\perp}- \pa_\a\pa_\m u_{\n\bet}^{\perp}\right\}\nn
&+& \tfrac{1}{8}\, \Big\{\eta_{\m\bet} \pa_\n\pa_\a + \eta_{\n\a} \pa_\m\pa_\bet
- \eta_{\m\a} \pa_\n\pa_\bet - \eta_{\n\bet} \pa_\m\pa_\a \Big\}(h-\sq w) 
\label{linRie}\\ 
\del G_{\m\n} &=&\del G_{\m\n}^{(T)} + \del G_{\m\n}^{(S)}=
 - \tfrac{1}{2} \sq h^{\perp}_{\m\n} + \tfrac{1}{4} (\eta_{\m\n} \sq - \pa_\m\pa_\n) (h - \sq w)
\label{linEinG}\\
\del R &=& -\tfrac{3}{4} \sq (h - \sq w) \label{delRicscal}
\eea
\label{linEinten}\ees
depending only upon the gauge invariant tensor $(T)$ and scalar $(S)$ parts of the linearized metric perturbations (\ref{gaugeinvcov})
respectively. Substituting the decomposition (\ref{metdecomp}) into (\ref{Seff2}) gives
\bea
&&S_{eff}^{(2)}\Big\vert_{g=\eta} = \sdfrac{1}{64 \pi G}  \int\, d^4x\, u^{\perp \m\n} \sq u_{\m\n}^{\perp}
-\sdfrac{3}{512 \pi G}  \int\, d^4x\, (h- \sq w)\sq (h - \sq w)\nn
&&\hspace{1.5cm} -\, \sdfrac{b'}{2\,}  \int\, d^4x\, (\sq\vf )^2 + \sdfrac{b'}{4\,}  \int\, d^4x\,  (\sq\vf )\sq(h- \sq w)
\label{Seff2inv}
\eea
which is also gauge invariant under linearized coordinate transformations. Thus the gauge invariant scalar metric
perturbations $h-\sq w$ couples to and mixes with the conformalon scalar via $\sq\vf$.

The tensor decomposition (\ref{metdecomp}) is useful for separating gauge invariant from gauge non-invariant
components of the linearized deviations of the metric from flat space in a Lorentz covariant way.
Because the Einstein equations obtained by the full metric variation has $tt$ and $ti$ components which act as
four constraints, similar to the Gauss Law constraint on the longitudinal part of the electric field in electromagnetism,
not all modes in the covariant decomposition (\ref{gaugeinvcov}) are true propagating degrees of freedom.
In the pure Einstein theory ($b'=0$) the scalar spin-$0$ metric perturbations $h- \sq w$ and $3$ of the $5$ 
polarizations of spin-2 gravitational waves are constrained, so that only the $2$ remaining transverse, traceless 
components freely propagate in classical GR.

The four first order constraints corresponding to the four diffeomorphism gauge degrees of freedom are the
$tt$ and $ti$ components of the linearized Einstein eqs. (\ref{Ein}). The scalar conformalon stress-energy tensor
(\ref{Eab}) in the flat space limit is
\bea
E_{\m\n}\big\vert_{flat} &=&  - 2\,(\na_{(\m}\vf) (\na_{\n)} \sq \vf)  + 2\, (\sq\vf)(\na_\m\na_\n\vf)
+ \tfrac{2}{3}\, (\na_\a\vf)(\na^\a\na_\m\na_\n\vf)
- \tfrac{4}{3}\,( \na_\m\na_\a \vf)(\na_\n\na^\a\vf)\nn
&&\ \quad +\, \tfrac{1}{6}\, \eta_{\m\n}\, \Big\{\!\!-3\, (\sq\vf)^2 + \sq \big[(\na_\a\vf)(\na^\a\vf)\big] \Big\}
+ \tfrac{2}{3}\,(\eta_{\m\n} \sq - \na_\m\na_\n)\,\sq \vf
\label{Eabflat}
\eea
with
\vspace{-5mm}
\be
T_{\m\n}[\vf]\big\vert_{flat} = b' E_{\m\n}\big\vert_{flat} \qquad {\rm and} \qquad C_{\m\n}\big\vert_{flat} = 0
\label{Cabflat}
\ee
{\it in vacuo} when no other sources are present. Note that (\ref{Eabflat}) contains terms both linear and quadratic in
$\vf$. The terms linear in $\vf$, (\ref{Ein}), and the scalar part of (\ref{linEinG}) give
\be
\del G_{\m\n}^{(S)} = \sdfrac{1}{4} (\eta_{\m\n} \sq - \pa_\m\pa_\n) (h - \sq w) = 
\sdfrac{16\pi G b'}{3}\,(\eta_{\m\n} \sq - \pa_\m\pa_\n)\,\sq \vf
\ee
or in component form,
\bes
\bea
\del G_{tt}^{(S)}= -\sdfrac{1}{4}\,\vec\na^2\, (h- \sq w)&=& -\sdfrac{16 \pi G b'}{3}\, \vec \na^2 (\sq \vf)\\
\del G_{ti}^{(S)} = -\sdfrac{1}{4}\,\pa_t\vec\na_i\, (h- \sq w) &=& -\sdfrac{16 \pi G b'}{3}\, \pa_t\vec\na_i (\sq \vf)
\eea
\label{LinEinvf}\ees
to linear order in $h_{\m\n}$ and $\vf$, with the spin-$2$ tensor part $\del G_{\m\n}^{(T)}$ dependent upon $u^{\perp}_{\m\n}$ unaffected.
The general solution of the constraints (\ref{LinEinvf}) in the scalar sector is
\be
\sdfrac{1}{8}\, (h - \sq w) = \sdfrac{8 \pi G b'}{3} \sq \vf + F(t) + K (\bx)\,,\qquad \vec\na^2 K = 0
\label{scaleq}
\ee
where $F(t)$ is a function only of time and $K(\bx)$ is a function only of space, with zero spatial Laplacian. Since the only solution
of $\vec\na^2 K = 0$ which is non-singular everywhere, including spatial infinity, is $K(\bx) = const.$, it may be absorbed
into $F(t)$, and we may effectively set $K=0$ in (\ref{scaleq}).

In the pure Einstein theory ({\it i.e.}\,$b'=0$) (\ref{scaleq}) implies that the gauge invariant scalar metric perturbation $h-\sq w = 8 F(t)$
contains no local propagating degree of freedom, in the same way that the longitudinal part of the electric field is locally constrained
by Gauss' Law $\vec\na \cdot \vec E = \rho$ and is non-propagating in pure electromagnetism (with $\rho =0$). In that case as well
the longitudinal electric field can contain at most a spatially homogeneous mode that is a function only of time: $\vec E = \vec E (t)$.
As is well known in finite temperature field theory, plasmas or generally whenever there is a polarizable fluctuating medium,
the fluctuations of the charges in the medium induce through Gauss' law a collective plasmon mode described by a propagating longitudinal
electric field \cite{LeB}. Thus the longitudinal mode of the electric field, completely constrained and non-propagating with fixed sources,
becomes propagating in the presence of dynamical charged sources, inheriting the collective dynamics of the sources.

An analogous phenomenon can occur due to quantum fluctuations and anomaly of a charged field in the vacuum, as evidenced
for example by the Schwinger model of massless QED in one $1\!+\!1$ dimensions. Integrating out of the massless fermion field leads
to a propagating massive scalar boson equivalent to the longitudinal electric field, classically constrained, which becomes propagating
through Gauss' Law and the quantum axial anomaly \cite{BlaCabMot}. In either case the collective mode
arises from two-particle correlations of the fluctuating source field (either thermal or quantum).

Eq. (\ref{scaleq}) shows that this phenomenon occurs {\it in vacuo} in the EFT of gravity due to the conformal trace anomaly,
since the gauge invariant scalar metric perturbations $h-\sq w$, previously constrained to be non-propagating by the classical Einstein eqs.
are forced now to follow the {\it propagating} conformalon scalar field $\sq \vf$. Since $\vf$ satisfies the fourth order eq.~(\ref{phieom}), or
\be
\sq^2 \vf = \sq (\sq \vf) = - \tfrac{1}{3}\, \sq \del R
\label{vfeomdelR}
\ee
linearized around flat space, it contains two sets of scalar field modes. In fact, as will be shown in the next subsection,
the linearized Einstein eqs.\,also imply $\del R =0$ identically at linearized order. Hence of two sets of massless modes
in the fourth order eq.~of motion for $\vf$, only the one remaining in $\sq \vf\neq 0$ survives to provide
\hspace{-7mm}
\be
\sq (h-\sq w) = 0
\label{scalGW}
\ee
with non-trivial propagating scalar gravitational wave solutions, obeying a {\it second order} wave equation.
In the next subsection a non-covariant space + time splitting of the metric perturbations, appropriate for identifying 
the unconstrained propagating local scalar degree of freedom in a canonical framework is given, in order to verify this conclusion.
\vspace{-3mm}

\subsection{Space + Time Decomposition and Constraints}

In order to properly characterize the dynamical propagating degrees of freedom in the EFT of gravity
{\it vs.} the constrained modes, introduce the non-covariant space + time splitting of the
linearized metric perturbations in the standard Hodge decomposition
\bes\bea
&&h_{tt}=-2 A \\
&&h_{ti}=\cB_i^{\perp} + \vec\na_i B  \\
&&\hspace{-3cm}h_{ij}=   \cH_{ij}^{\perp} + \vec\na_i\cE_j^{\perp} + \vec\na_j\cE_i^{\perp} + 2\,\eta_{ij}\, C +
2\,\big( \vec\na_i\vec\na_j - \tfrac{1}{3}\,\eta_{ij} \vec\na^2 \big)D
\eea
\label{linmet}\ees
adapted from that commonly employed in spatially homogeneous, isotropic cosmological models \cite{Bard,DSAnom},
with $\eta_{ij} = \del_{ij}$ in Cartesian coordinates of flat $\R^3$. The components $A, B, C, D$ are four scalars with respect
to the flat spatial metric on $\R^3$, the $3$-vectors $\cB_i^{\perp}, \cE_i^{\perp}$ are transverse, satisfying
\be
\vec\na_i \cB_i^{\perp} = \vec\na_i \cE_i^{\perp} = 0
\ee
resulting in two independent components each, and $\cH_{ij}^{\perp}=  \cH_{ji}^{\perp}$ is a symmetric, transverse,
traceless tensor (now in the spatial three-dimensional sense) satisfying
\be
\vec\na_j \cH_{ij}^{\perp} = \vec\na_i  \cH_{ij}^{\perp}  = 0= \eta^{ij} \, \cH_{ij}^{\perp}
\ee
resulting in two independent polarization components.  

In this space + time splitting the four independent linearized coordinate transformations can be similarly decomposed:
\vspace{-5mm}
\be
\xi_t = -T\hspace{4cm} \xi_i = \vec\na_i L + \cX_i^{\perp}
\ee
with $\cX_i^{\perp}$ a transverse vector satisfying $\vec\na_i\cX_i^{\perp} = 0$. The linearized
coordinate gauge transformation induces the linearized changes in the metric components
\be
\begin{aligned}[c]
A &\ra A + \dt{T}\\
B &\ra  B - T + \dt{L}\\
C &\ra  C + \tfrac{1}{3}\vec\na^2 L\\
D &\ra  D + L
\end{aligned}
\hspace{4cm}
\begin{aligned}[c]
\cB_i^{\perp} &\ra \cB_i^{\perp} + \dt{\cX}_i^{\perp}\\
\cE_i^{\perp} &\ra \cE_i^{\perp} + \cX_i^{\perp}\\
\cH_{ij}^{\perp} & \ra  \cH_{ij}^{\perp}
\end{aligned}
\label{noncovgauge}
\ee
provided the spatial gradients of $B, D, T$ and $L$ are all non-vanishing. Thus
\be
\begin{aligned}[c]
\Y_A &\equiv A + \dt{B} - \ddt{D}\\
\Y_C &\equiv C - \tfrac{1}{3} \vec\na^2 D
\end{aligned}
\hspace{4cm}
\begin{aligned}[c]
\Psi_i^{\perp} &\equiv \cB_i^{\perp}- \dt{\cE}_i^{\perp}\\
\ {\rm and} &\quad  \cH_{ij}^{\perp}
\end{aligned}
\label{noncovgaugeinv}\ee
are invariant under linearized coordinate transformations, whenever the spatial dependences
of these functions are non-vanishing. In the special case of metric perturbations independent of the spatial
coordinate $\bf x$, $B, D$ and $L$ may be set to zero. Then only $\Y_C = C$ is gauge invariant, while $\Y_A = A$
is still subject to the gauge transformation, $\Y_A \ra\, A \,+\, \dt{T}$, with $T(t)$ independent of the spatial coordinate.

The non-zero components of the Riemann tensor can be expressed in the form
\vspace{-3mm}
\bes
\begin{align}
\del R_{titj} &= \pa_i\pa_j \Y_A - \eta_{ij} \ddt{\Y}_C + \tfrac{1}{2} \left(\pa_i\Psi_j^{\perp} + \pa_j \Psi_i^{\perp}\right) - \tfrac{1}{2} \ddt{\cH}_{ij}^{\perp}\\
\del R_{tijk} &= \left(\eta_{ij} \pa_k - \eta_{ik} \pa_j\right) \dt{\Y}_C + \tfrac{1}{2}\, \pa_i \left( \pa_j\Psi_k^{\perp} - \pa_k \Psi_j^{\perp}\right)
+ \tfrac{1}{2} \left(\pa_k \dt{\cH}_{ij}^{\perp} - \pa_j \dt{\cH}_{ik}^{\perp}\right)\\
\del R_{ijkl} &= \big(\eta_{il}\, \pa_j\pa_k - \eta_{jl}\, \pa_i\pa_k + \eta_{jk}\, \pa_i\pa_l  - \eta_{ik}\, \pa_j\pa_l\big)\Y_C\nn
& \quad + \tfrac{1}{2} \,\big( \pa_k\pa_j\cH_{il}^{\perp} - \pa_k\pa_i \cH_{jl}^{\perp} - \pa_j\pa_l \cH_{ik}^{\perp}  + \pa_i\pa_l \cH_{jk}^{\perp} \big)
\end{align}
\vspace{-3mm}
\label{varRie}\ees
from which follow the linearized variation of the Ricci tensor and Ricci scalar
\bes
\bea
\hspace{1cm}&&\del R_{tt} = \vec\na^2\Y_A - 3\, \ddt{\Y}_C\\
&&\del R_{ti} = - 2\, \pa_i \dt{\Y}_C - \tfrac{1}{2} \vec\na^2 \Psi_i^{\perp} \\
&&\del R_{ij} = -\pa_i\pa_j \Y_A - (\pa_i\pa_j + \eta_{ij} \vec\na^2) \Y_C + \eta_{ij} \ddt{\Y}_C
- \tfrac{1}{2} (\pa_i \Psi_j^{\perp} + \pa_j \Psi_i^{\perp}) -\tfrac{1}{2} \sq \cH_{ij}^{\perp}
\label{varRic}\\
&&\del R = -2\, \vec\na^2 \Y_A  - 4\, \vec\na^2 \Y_C + 6\, \ddt{\Y}_C 
\label{varRicscal}
\eea
\ees
and the linearized Einstein tensor
\bes
\begin{align}
\del G_{tt} &= -2\, \vec\na^2\Y_C\label{UpsCeq}\\
\del G_{ti} &=- 2\, \pa_i \dt{\Y}_C - \tfrac{1}{2} \vec\na^2 \Psi_i^{\perp}\\
\del G_{ij} &= (\eta_{ij} \vec\na^2 -\pa_i\pa_j) (\Y_A + \Y_C)  -2\,\eta_{ij} \ddt{\Y}_C
- \tfrac{1}{2} (\pa_i \Psi_j^{\perp} + \pa_j \Psi_i^{\perp}) -\tfrac{1}{2} \sq \cH_{ij}^{\perp} \label{UpsAeq}
\end{align}
\label{varEin}\ees
all expressed in terms of the six gauge invariant components (\ref{noncovgaugeinv}).
From (\ref{UpsCeq}) it is clear that the $tt$ component of the Einstein's eqs.,
$\del G_{tt} = 8 \pi G \,\del T_{tt}$ is
\be
\vec\na^2 \Y_C = - 4 \pi G\, \del T_{tt}
\ee
for small perturbations about flat space.  Since $\del G_{ij} \ra 0$ in the non-relativistic limit for slowly moving weak
sources, $\Y_A = -\Y_C$ from (\ref{UpsAeq}) becomes the Newtonian potential for quasi-static weak sources.
On the other hand for relativistic sources, $\del G_{ij} \neq 0$ and $\Y_A \neq - \Y_C$ in general.

The quadratic effective action $S^{(2)}_{eff}$ around flat space in the decomposition (\ref{linmet}) takes the form
\bea
&&S^{(2)}_{eff}\Big\vert_{g=\eta} = \sdfrac{1}{64\pi G}\! \int d^4x\! \left(\cH_{ij}^{\perp}\sq \cH_{ij}^{\perp} - 2 \Psi_i^{\perp} \vec\na^2 \Psi_i^{\perp}\right)
+ \sdfrac{1}{8 \pi G}\! \int \!d^4 x \left( -2 \Y_A \vec\na^2 \Y_C - \Y_C\vec\na^2 \Y_C + 3 \Y_C\ddt{\Y}_C\right)\nn
&&\hspace{2cm}- \sdfrac{b'}{2} \int d^4 x \,(\sq \vf)^2 - \sdfrac{2b'}{3} \int d^4x\, (\sq \vf) \left(-\vec\na^2 \Y_A + 3\, \ddt{\Y}_C - 2\, \vec\na^2 \Y_C\right)
\label{SeffUpsAC}
\eea
while the linearized Einstein equations become
\bes
\bea
\hspace{3cm}\sq \cH_{ij}^{\perp} &=&0\label{TT}\\
\vec\na^2 \Psi_i^{\perp} &=&0\label{vecT}\\
\vec\na^2 \Y_C &=&\sdfrac{8 \pi G b'}{3} \, \vec\na^2 (\sq \vf)\label{Eintt} \\
\pa_t \vec\na_i\Y_C &=& \sdfrac{8 \pi G b'}{3} \, \pa_t\vec\na_i(\sq \vf)\label{Einti}\\
&&\hspace{-5cm}(\eta_{ij} \vec\na^2 - \vec\na_i\vec\na_j)( \Y_A + \Y_C) - 2 \eta_{ij}\, \ddt{\Y}_C =  -\sdfrac{16 \pi G b'}{3}
 \, \vec\na_i\vec\na_j (\sq \vf)
\label{Einij}
\eea
\label{linEin}\ees
since the linearly independent components in the tensor, vector and scalar sectors must be satisfied separately.
Note that although the variation of the scalar effective action (\ref{Seff2}) yields the scalar solutions of the linearized
Einstein equations (\ref{linEin}), the latter tensorial equations are more restrictive, and it is only this more restrictive set
of solutions of the tensorial eqs.~(\ref{linEin}) that satisfy all the correct linearized constraints of diffeomorphism invariance.
The first eq.~(\ref{TT}) gives the usual three-dimensionally transverse, traceless, propagating gravitational wave modes,
while the second eq.~(\ref{vecT}) expresses the triviality of the transverse, vector sector in the absence of
any rotating sources. The last three eqs.~of (\ref{linEin}) together with (\ref{vfeomdelR}) show the non-trivial coupling of
the anomaly scalar into the scalar sector of metric perturbations in the space + time splitting.

Analogously to the covariant analysis in the previous subsection, the constraint eqs. (\ref{Eintt}) and (\ref{Einti}) imply
\vspace{-3mm}
\be
\Y_C = \sdfrac{8 \pi G b'}{3} \, \sq \vf + F(t)
\label{UpsCsoln}
\ee
with $F$ some function only of time. When twice (\ref{Eintt}) is added to the trace of (\ref{Einij}) we obtain
\be
 2\,\vec\na^2  \Y_A +4\, \vec\na^2 \Y_C  -6\, \ddt{\Y}_C = -\del R = 0
\label{delRzero}
\ee
by (\ref{varRicscal}), showing that the Ricci curvature scalar perturbations are constrained to be identically zero at linearized order
as a direct consequence of the linearized Einstein eqs.~(\ref{linEin}). As anticipated this then implies that the $\vf$ equation
of motion (\ref{vfeomdelR}) linearized around flat space becomes
\be
\sq^2 \,\vf = -\tfrac{1}{3} \,\sq \del R =  0\,.
\label{vfeomvac}
\ee
which is sourcefree {\it in vacuo}. Finally (\ref{delRzero}) together with (\ref{UpsCsoln}) and (\ref{vfeomvac}) imply that
\be
\vec\na^2 (\Y_A-\Y_C) = 3\, ( \ddt{\Y}_C - \vec\na^2 \Y_C) = - 8\pi G b' \sq^2 \vf + 3 \ddt{F} = 3 \ddt{F}\,.
\label{UpsACF}
\ee
Since the right side of this relation is a constant in space, but the spatial Laplacian $\vec\na^2$ has no non-singular
inverse on constant functions, the only non-singular solutions of (\ref{UpsACF}) are those where each of
the four different expressions in (\ref{UpsACF}) are all vanishing. Thus in addition to (\ref{vfeomvac}), we find
\bes
\bea
\ddt{F}(t) &=& 0\label{Fsoln}\\
\Y_A - \Y_C &=& \dt{\xi}(t)
\eea
\ees
where $\dt{\xi}(t)$ is another arbitrary function of time. This arbitrary function $\dt{\xi}(t)$ may be removed by the residual time coordinate
gauge freedom $\Y_A \ra \Y_A + \dt{T}$ allowed for spatially homogeneous time reparameterizations, so that
choosing $T(t) = - \xi(t)$ fixes
\be
\Y_A = \Y_C =  \sdfrac{8 \pi G b'}{3} \, \sq \vf + F(t) = \sdfrac{1}{8} \,(h-\sq w)
\label{UpsACsoln}
\ee
with $F(t) = \alpha t + \beta$ at most a linear function of time due to (\ref{Fsoln}), and the last equality follows from (\ref{scaleq})
with $K=0$ with $F(t)$ identified as the same function as in the covariant analysis, {\it cf.} (\ref{scaleq}).

Thus the solution of the Einstein eqs., the $\vf$ equation of motion (\ref{vfeomvac}) and all diffeomorphism constraints
for the gauge invariant scalar potentials imply finally 
\be
\sq \Y_A = \sq \Y_C = \sdfrac{8 \pi G b'}{3} \, \sq^2 \vf = 0
\label{scalpotwvs}
\ee
describing true propagating scalar gravitational waves in flat space, arising from the anomaly scalar for $\sq \vf \neq 0$.
The equal gravitational potentials $\Y_A = \Y_C$ (up to a pure gauge) with (\ref{UpsACsoln}) make clear that the gauge invariant and
Lorentz invariant scalar metric perturbation $h- \sq w$, constrained in the classical Einstein theory is a {\it bona fide}
propagating scalar wave mode in the semi-classical EFT. Note that this discontinuous change in the nature of
the scalar sector of GR when $b'\neq 0$ is unrelated to the Planck scale and hence unsuppressed at
low energies and macroscopic distance scales, and moreover, half of the solutions of (\ref{vfeomvac}) for which $\sq \vf =0$
decouple from the metric perturbations at linear order around flat space. 

For completeness we record the relations between the covariant metric decomposition (\ref{metdecomp}) and 
the $3+1$ decomposition (\ref{linmet}):
\bes
\bea
u_{tt}^\perp &=& - \frac{4}{3} \frac{1\ }{\sq^2} (\vec \na^2)^2\, (\Y_A -\Y_C)\\
u_{ti}^\perp &=& \frac{1}{\sq} \vec\na^2 \Psi_i^\perp - \frac{4}{3} \frac{1\,}{\sq^2} \vec\na_i \vec\na^2\, (\dt\Y_A - \dt\Y_C)\\
u_{ij}^\perp &=& \cH_{ij}^\perp + \frac{1}{\sq} \left(\vec\na_i \dt\Psi^\perp_j + \vec\na_j \dt\Psi^\perp_i\right)
- \frac{2\ }{\sq^2} \vec\na_i\vec\na_j \left(\ddt\Y_A - \ddt\Y_C\right) \nn
&&\hspace{3cm}+ \frac{2}{3}\frac{1\ }{\sq^2} \left(\vec\na_i\vec\na_j  - \eta_{ij} \sq\right)\vec\na^2\, (\Y_A- \Y_C) \nn
h- \sq w &=& 8\, \Y_C + \frac{8}{3} \frac{1}{\sq}\, \vec\na^2\, (\Y_A - \Y_C)
\eea
\ees
which show that the scalar wave solution (\ref{UpsACsoln}) with $\Y_A= \Y_C$ and $\Psi_i^\perp = 0$ is pure gauge invaraint spacetime 
scalar $h- \sq w$, and leaves the transverse, traceless gravitational wave sector of the classical Einstein theory, with its $2$ gauge invariant 
propagating polarizations $u^{\perp}_{ij} = \cH^{\perp}_{ij}$ unaffected.

\section{Scalar Gravitational Waves from Localized Sources}
\label{Sec:Green}

Having found that the semi-classical EFT of gravity admits scalar gravitational wave solutions {\it in vacuo} for any $b' \neq 0$, which
are not present in classical GR, in this section we consider localized sources for these scalar waves, arising from the anomaly stress tensor,
and solve the linearized equations with fixed causal (classical retarded) boundary conditions due to these localized sources.

In flat spacetime the eq.~of motion (\ref{phieom}) for the anomaly scalar field $\vf$ is
\be
\sq^2\vf = 8 \pi J
\label{eomflat}
\ee
with the source
\vspace{-2mm}
\be
J \equiv \frac{1}{16\pi b'}\,\frac{\cA}{\sqrt{-g}} =\frac{1}{16 \pi} \Big[E - \tfrac{2}{3}\sq R + \frac{b\,}{b'}\, C^2 + \frac{1}{b'}\sum_i \bet_i\, \cL_i\Big]
\rightarrow \frac{1}{16\pi b'}\sum_i \bet_i\, \cL_i
\label{defJ}
\ee
to be treated as a weak perturbation in the nearly flat space limit. The factor of $8 \pi$ is inserted in (\ref{eomflat})
for later convenience. The linear equation (\ref{eomflat}) can be solved in terms of the classical retarded Green's function
$(\sq^{-2})_{Ret} \equiv D_{R}$, which in flat spacetime (dropping the subscript $4$ on $D_4$) satisfies
\vspace{-3mm}
\be
\sq^2 D_{R}(t-t'; \bx - \bxp) = \del (t-t')\,\del^3(\bx -\bxp)
\vspace{-3mm}
\ee
and which is represented by the Fourier integral
\be
D_R(t-t'; \bx - \bxp) = \int_{-\infty}^{\infty}\frac{d \w}{2 \pi} \int \frac{d^3 \bk}{(2\pi)^3}\ \frac{e^{-i \w (t-t')}\, e^{i \bk\cdot (\bx-\bxp)}}
{\big[-(\w +i \e)^2 + \bk^2\big]^2}
\ee
with the infinitesimal $\e > 0$ prescription enforcing retarded boundary conditions. Since the second order poles of the integrand
at $\w = \pm |\bk|$ are displaced into the lower half complex by this prescription, the $\w$ contour may be
closed in the upper half complex plane for $t-t' < 0$, giving zero, while for $t-t' > 0$ the $\w$ contour may
be closed in the lower half complex plane, yielding the result
\bea
&&D_R(t-t'; \bx - \bxp)=  \int \frac{d^3 \bk}{(2\pi)^3}\ \frac{e^{i \bk\cdot (\bx-\bxp)}} {2\, k^2}\,
\left\{ \frac{\sin\left[k (t-t')\right]}{k} - (t-t') \,\cos\left[k\,(t-t')\right] \right\}\,\theta (t-t')\nn
&& \hspace{1cm}=\frac{1}{4 \pi^2} \int_0^{\infty} dk\ \frac{\sin\!\big[k \,|\bx-\bxp|\big]}{k\, |\bx-\bxp|} \,\left\{ \frac{\sin\left[k\,(t-t')\right]}{k}
- (t-t') \,\cos\left[k\,(t-t')\right]\right\} \,\theta (t-t')\nn
&&\hspace{3cm}= \frac{1}{8 \pi} \ \theta \big(t-t' -  |\bx-\bxp|\big) \quad{\rm for}\ \ t > t'  \qquad (0\ \  \rm otherwise)
\label{DR}
\eea
which vanishes if $t\le t'$, and where eqs. 3.741 (2) and (3) of  \cite{GraRyz} and have been used.

Thus whereas the usual second order wave operator $\sq$ has a retarded Green's function
\be
\frac{1}{4 \pi |\bx-\bxp|} \ \del \big(t-t' - |\bx-\bxp|\big)\, \theta (t-t') = -\sq D_R(t-t'; \bx - \bxp)
\ee
with support only on the past light cone in $3+1$ spacetime dimensions, the fourth order operator $\sq^2$
has a retarded Green's function (\ref{DR}) with uniformly constant support everywhere within the past light cone.
This is similar to the retarded Green's function for the second order wave operator $\sq$ but in $1 +1$ spacetime dimensions,
and results in pronounced long distance infrared behavior of the conformalon scalar $\vf$. Indeed the solution
of (\ref{eomflat}) fixed by these classical retarded boundary conditions is
\bea
\vf (t, \bx) &=& 8 \pi \int_{-\infty}^{\infty} dt' \int d^3 \bxp\, D_R(t-t'; \bx - \bxp) \,J(t', \bxp) \nn
&=&  \int d^3 \bxp  \int_{-\infty}^{t -  |\bx-\bxp|}  \! dt' \,J(t', \bxp)
\label{vfsol}
\eea
which does not fall off with large $ |\bx-\bxp|$, and which can even become large without bound for persistent sources.
Inspection of (\ref{Eab}) shows that $\vf$ appears only under derivatives, so that these largest infrared effects
are removed for localized sources.

Relabeling the point of observation $\bx \ra \bR$, and the integration variable over
the spatial extent of the source $\bxp \ra \bx$, the first derivatives of (\ref{vfsol}) are
\bes
\bea
\dt{\vf} (t, \bR) &=&  \int d^3 \bx \,J(\tilt, \bx)\\
\vec\na_i\, \vf (t, \bR) &=& -\int d^3 \bx\ \bn_i \ J(\tilt, \bx)
\eea
\label{fdvf}
\ees
where
\be
\bn_i \equiv \vec\na_i\, |\bR -\bx| = \frac{(\bR-\bx)_i}{|\bR-\bx|\,}\qquad {\rm and} \qquad \tilt \equiv t - |\bR-\bx|
\ee
is the retarded time, the spatial derivatives being taken with respect to $\bR$. The second derivatives of  (\ref{vfsol}) are
\bes
\bea
\ddt{\vf} (t, \bR) &=&  \int d^3 \bx \,\dt{J}(\tilt, \bx)\\
\vec\na_i\, \dt{\vf} (t, \bR) &=&  -\int d^3\bx\  \bn_i \,\dt{J}(\tilt, \bx)\\
\vec\na_i\vec\na_j\, \vf (t, \bR) &=&  -\int d^3 \bx\ \frac{ \big(\del_{ij} - \bn_i\bn_j\big)}{|\bR-\bx|\,}\,J(\tilt, \bx)
+ \int d^3\bx\  \bn_i \bn_j\,\dt{J}(\tilt, \bx)
\eea
\label{sdvf}
\ees
from which follow
\vspace{-3mm}
\bes
\bea
\vec\na^2\vf (t, \bR) &=&  -2 \int d^3 \bx\ \frac{1}{|\bR-\bx|\,}\,J(\tilt, \bx) + \int d^3\bx\, \dt{J}(\tilt, \bx)\\
\sq\vf (t, \bR) &=& ( -\pa_t^2  + \vec\na^2 )\,\vf (t, \bR) = -2 \int d^3 \bx\ \frac{1}{|\bR-\bx|\,}\, J(\tilt, \bx)\\
\sq\, \dt{\vf}(t, \bR) &=&  -2 \int d^3 \bx\ \frac{1}{|\bR-\bx|\,}\, \dt{J}(\tilt, \bx)\,.
\eea
\label{sdsqvf}\ees
Thus from (\ref{UpsACsoln}) the gauge invariant scalar metric perturbation propagated from a distant localized source is
\vspace{-3mm}
\bea
\Y_A = \Y_C = \frac{1}{8} (h- \sq w) &=& -\frac{16 \pi G b'}{3} \int d^3 \bx\ \frac{1}{|\bR-\bx|\,}\, J(\tilt, \bx)\nn
&\rightarrow&  -\frac{G}{3r} \,\int d^3 \bx\,\cA (\tilt, \bx)
\label{scalpert}
\eea
in the far or radiation zone where $r \equiv |\bR| \gg |\bx|$.

For time harmonic sources,
\be
\cA (t, \bx) = e^{-i \w t} \cA_\w (\bx)
\label{timeosc}
\ee
where the Real Part is understood, we have
\be
\int d^3 \bx\, \cA(\tilt,\bx) \simeq e^{-i\w ( t-r) }\,  \int d^3 \bx \, \exp (-i \w \br \cdot \bx)\,\cA_\w (\bx) \equiv e^{-i\w ( t-r) }\, \tilde\cA (\w|\br)
\label{Iomega}
\ee
in the far radiation zone, and therefore from (\ref{scalpert}),
\be
\Y_A = \Y_C = - \frac{G}{3r}\, e^{-i\w ( t-r) } \, \int d^3 \bx \, \exp (-i \w \br \cdot \bx)\,\cA_\w (\bx)  = - \frac{G}{3r}\, e^{-i\w ( t-r) } \,\cA (\w|\br)
\label{scalgwave}
\ee
which has the form of an outgoing spherical scalar gravitational wave.

This result is to be compared with the linearized metric perturbation of transverse, traceless gravitational waves
\vspace{-5mm}
\bea
\cH_{ij}^{\perp}(t, \bR) &=& 4G\, \int d^3\bx \, \frac{1}{|\bR - \bx|} \ T_{ij}^{\perp} (\tilt, \bx)\nn
&\ra&\frac{4G}{r} \,\Pi_{ij}^{\ \ lm}(\br) \int d^3\bx \, T_{lm} (t -r + \br\cdot \bx, \bx)\nn
&=&\frac{2G}{r}\, \Pi_{ij}^{\ \ lm}(\br)\int d^3\bx\, \bx_l\,\bx_m\,\ddt{T}^{tt}(t-r + \br\cdot \bx, \bx)
\label{TTGW}
\vspace{-4mm}
\eea
in the radiation zone, where
\vspace{-2mm}
\bea
&&\Pi_{ij}^{\ \ lm}(\br) = \sdfrac{1}{2} \left(\delta_i^{\ l} \delta_j^{\ m} + \delta_i^{\ m} \delta_j^{\ l} 
- \eta_{ij}\eta^{lm} +  \eta_{ij}\, \br^l\,\br^m + \br_i\,\br_j\,\eta^{lm} -\,\delta_i^{\ l}\, \br_j\,\br^m - \delta_j^{\ l}\, \br_i\,\br^m\right.\nn
&&\hspace{4cm}\left.  -\, \delta_i^{\ m}\, \br_j\,\br^l -\delta_j^{\ m}\, \br_i\,\br^l 
+ \br_i\,\br_j\,\br^l\,\br^m\right)
\vspace{-3mm}
\eea
is the projector onto transverse, traceless tensors. For a time harmonic source
\bea
\cH_{ij}^{\perp}(t, \bR)&=&   -\frac{2G\w^2}{r}\, e^{-i\w(t-r)}\,\Pi_{ij}^{\ \ lm}(\br)\int d^3\bx \, \exp( -i\w\br\cdot \bx)\, \bx_l\,\bx_m\,
T_\w^{tt} (\bx)\nn
&\simeq& -\frac{2G\w^2}{r}\, e^{-i\w(t-r)}\,\Pi_{ij}^{\ \ lm}(\br)\ \mQ_{lm}(\w)
\label{TTQuad}
\eea
in the quadrupole approximation, where the exponential factor in the integral in (\ref{TTQuad}) is replaced by unity,
valid if $\w a/c \ll 1$ for slow moving non-relativistic sources of spatial extent $a$, and where
\be
\mQ_{lm}(\w) = \int d^3\bx \,\bx_l\,\bx_m\,T_\w^{tt} (\bx)
\label{Quadmom}
\ee
is the Fourier component of the source quadrupole moment.

Thus whereas the source for electromagnetic radiation is the time varying current transverse to the line of sight,
whose first non-vanishing multipole is the dipole term, and the source for transverse, traceless gravitational
radiation in classical GR is the time varying transverse stress-energy whose first non-vanishing multipole is the
quadrupole term (\ref{Quadmom}), the source for the scalar gravitational wave in (\ref{scalpert})-(\ref{scalgwave}) 
is the trace anomaly $\cA$, which receives a contribution from its monopole term in the expansion of the exponent 
in (\ref{scalgwave}). Due to the decoupling of the anomaly gauge field sources when $\w_k < m_e$ (QED) or 
$\w_k < m_{u,d}$ (QCD) an effective lower threshold for space and time variation of these sources is introduced 
for scalar monopole radiation. As shown below and in the next section, the Riemann tensor and
and power radiated in scalar gravitational radiation are proportional to $\w^2$.

The linearized perturbation of the Riemann tensor corresponding to the scalar gravitational wave (\ref{scalgwave})
may be computed with the help of (\ref{linRie}) to be
\bea
\del R_{\m\n\a\bet}^{(\cA)} &=&
\Big\{\eta_{\m\a} \pa_\n\pa_\bet + \eta_{\n\bet} \pa_\m\pa_\a -\eta_{\m\bet} \pa_\n\pa_\a - \eta_{\n\a} \pa_\m\pa_\bet\Big\}\,
\frac{G}{3\, r}\int d^3 \bx\,\cA(\tilt, \bx)\nn
&\ra & \frac{G}{3\, r}\, \Big\{\eta_{\m\a} \pa_\n\pa_\bet + \eta_{\n\bet} \pa_\m\pa_\a -\eta_{\m\bet} \pa_\n\pa_\a - \eta_{\n\a} \pa_\m\pa_\bet\Big\}
\int d^3 \bx\,\cA(\tilt, \bx)
\label{delRieS}
\eea
in the far region where the terms dropped by neglecting the effect of the derivatives upon $1/r$ fall off faster than $1/r$
as $r \ra \infty$. In this region because of the dependence of the integrand on the retarded time $\tilt = t - |\bR - \bx|$,
one can also make the replacement $\vec \na_i \ra - \br_i\, \pa_t$. From (\ref{delRieS}) 
the components of the Riemann tensor perturbation then become
\bes
\bea
\del R_{itjt}^{(\cA)} &=&\frac{G}{3\, r}\, \big(\eta_{ij} - \br_i\br_j\big)\int d^3 \bx\,\ddt{\cA}(\tilt, \bx)\\
\del R_{ijkt}^{(\cA)} &=&\frac{G}{3\, r}\, \big(\eta_{ik}\br_j - \eta_{jk}\br_i\big)\int d^3 \bx\,\ddt{\cA}(\tilt, \bx)\\
\del R_{ijkl}^{(\cA)} &=&\frac{G}{3\, r}\, \big(\eta_{ik} \br_j\br_l + \eta_{jl} \br_i\br_k -\eta_{il} \br_j\br_k - \eta_{jk} \br_i\br_l\big)
\int d^3 \bx\,\ddt{\cA}(\tilt, \bx)
\eea
\label{Riescal}\ees
in the far field region, whereas the linearized Weyl tensor
\be
\del C^{(\cA)}_{\m\n\a\bet} = 0
\label{Weyl0}
\ee
vanishes for scalar gravitational waves. The Riemann tensor perturbations (\ref{Riescal}) may be used to compute
the effect of the scalar gravitational wave on test masses in a detector. The contractions of (\ref{Riescal}) are
\vspace{-5mm}
\bes
\bea
\del R_{tt}^{(\cA)} &=&\frac{2G}{3\, r}\, \int d^3 \bx\,\ddt{\cA}(\tilt, \bx)\\
\del R_{it}^{(\cA)} &=&\frac{2G}{3\, r}\, \br_i \int d^3 \bx\,\ddt{\cA}(\tilt, \bx)\\
\del R_{ij}^{(\cA)} &=&\frac{2G}{3\, r}\, \br_i\br_j \int d^3 \bx\,\ddt{\cA}(\tilt, \bx)\\
\del R^{(\cA)} &=&0
\eea
\label{Ricscal}\ees
and for time harmonic sources (\ref{timeosc}) in the radiation zone we may make the replacement
\be
\int d^3 \bx\,\ddt{\cA}(\tilt, \bx) \ra  - \w^2\, e^{-i\w(t-r)} \, \tilde\cA(\w|\br)
\ee
in (\ref{Riescal}) and (\ref{Ricscal}), with $\tilde\cA(\w |\br)$ defined by (\ref{Iomega}), and the Real Part is understood.

Whereas the tensor perturbations $\cH_{ij}^{\perp}$ have vanishing linearized Ricci tensor, but
non-vanishing linearized Weyl tensor, befitting a spin-$2$ tensor gravitational wave, the
scalar gravitational waves $\Y_A=\Y_C$ generated by the trace anomaly have non-vanishing Ricci
tensor (\ref{Ricscal}), but vanishing Weyl tensor (\ref{Weyl0}). Both scalar and tensor waves fall off as 
$1/r$ from a distant source, with the anomaly monopolar source $\cA$ in (\ref{scalpert}) 
taking the place of the transverse, traceless tensor source $T_{ij}^{\perp}$ in (\ref{TTGW}).

\section{Energy and Power Radiated in Scalar Gravitational Waves}
\label{Sec:EnerFlux}

The terms linear in $\vf$ and the metric perturbation around flat space $h_{\m\n}$ in the EFT give rise to
scalar gravitational waves. As in the case of tensor perturbations in Einstein's theory, it is necessary
to consider the quadratic terms in the expansion around flat space in order to compute the energy
and power radiated by these waves. These quadratic terms arise from (\ref{Seff}) in
three possible ways:
\vspace{-2mm}
\begin{enumerate}
\item Quadratic terms in $h_{\m\n}$ in the Einstein-Hilbert action and effective stress tensor;
\vspace{-3mm}
\item Mixed terms linear in each of $h_{\m\n}$ and $\vf$ in the anomaly action and stress tensor;
\vspace{-3mm}
\item Quadratic terms in $\vf$, but lowest order in the metric in the anomaly action and stress tensor.
\end{enumerate} 
\vspace{-5mm}

\subsection{Einstein-Hilbert Quadratic Terms}

The terms of the first kind are of the same origin as in classical General Relativity (albeit in the scalar sector), 
and are encapsulated in the expansion of the Ricci tensor 
\be
R^{(2)}_{\m\n} = \na_{\a}\G^{(2)\a}_{\ \quad\m\n}-  \na_{\n}\G^{(2)\a}_{\ \quad\m\a} + \G^{(1)\a}_{\ \quad\m\n} \, \G^{(1)\bet}_{\ \quad\a\bet}
- \G^{(1)\a}_{\ \quad\m\bet}\, \G^{(1)\bet}_{\ \quad\a\n}
\label{Rictensec}
\ee
to second order in the metric perturbation (\ref{hdef}). Here $\G^{(\ell)\a}_{\ \quad\m\n}$ is the Christoffel connection at order $\ell$.
In accordance with the Brill-Hartle-Isaacson averaging procedure for gravitational waves of small amplitude and wavelength much 
shorter than any background curvature radius \cite{BriHarIsac}, one can ignore total derivatives in the stress tensor
such as the first two terms in (\ref{Rictensec}). Substituting the expression
\be
\G^{(1)\a}_{\ \quad\m\n} = \sdfrac{1}{2} \left( \pa_{\n} h^{\a}_{\ \m} +  \pa_{\m} h^{\a}_{\ \m} -  \pa^{\a} h^{\m \n} \right)
\ee
for the expansion of the Christoffel connection to the first order into (\ref{Rictensec}) one obtains
\bes
\bea
\lag R^{(2)}_{\m\n}\rag &=&\left\lag \sdfrac{1}{2} (\pa_{(\m}h) (\pa_\a h^{\a}_{\ \n)}) - \sdfrac{1}{4} (\pa_{\a}h) (\pa^\a h_{\m\n}) 
+ \sdfrac{1}{2} (\pa^\bet h^{\a}_{\ \m} ) (\pa_\bet h_{\n\a})\right.\nn
&&\hspace{2cm}\left. - \sdfrac{1}{2} (\pa_\a h^{\a}_{\ \m} ) (\pa_\bet h^{\bet}_{\ \n}) - \sdfrac{1}{4} (\pa_{\m}h^{\a\bet}) (\pa_\n h_{\a\bet}) \right\rag\\
\lag R^{(2)}\rag &=&  \left\lag -h^{\m\n} R^{(1)}_{\m\n} + \eta^{\m\n} R^{(2)}_{\m\n}\right\rag\nn
&=&\left\lag\sdfrac{1}{2} (\pa_\a h^{\a}_{\ \lam} ) (\pa_\bet h^{\bet\lam})
- \sdfrac{1}{4} (\pa_{\lam}h^{\a\bet}) (\pa^\lam h_{\a\bet}) - \sdfrac{1}{4} (\pa_{\a}h) (\pa^\a h) \right\rag
\eea
\ees
where all indices are raised and lowered with the flat space Minkowski metric, $h \equiv \eta^{\m\n}h_{\m\n}$,
integration by parts has been used freely under the averaging brackets, and $R^{(1)}_{\m\n} \equiv \del R_{\m\n}$ to first order
in the metric perturbation is given by (\ref{linEin}) with $R^{(1)} \equiv \del R = 0$ by (\ref{vfeomvac}).

The contribution of this first ({\bf A}) set of terms to the effective energy-momentum tensor of gravitational waves is 
the negative of the second order Einstein tensor and thus given by
\vspace{-2mm}
\be
- \sdfrac{1}{8 \pi G}\, \left\lag G^{(2)}_{\m\n} \right\rag = - \sdfrac{1}{8 \pi G}\, \left\lag R^{(2)}_{\m\n} - \sdfrac{1}{2}  \eta_{\m\n}R^{(2)} \right\rag 
= - \sdfrac{1}{8 \pi G}\, \left\lag R^{(2)}_{\m\n}\right\rag
\label{GW1}
\ee
where the second equality follows by substituting the covariant metric decomposition (\ref{metdecomp}), 
retaining only the $w$ and $h$ terms for scalar perturbations, so that
\bea
\vspace{-6mm}
\lag R^{(2)}\rag &=& -\sdfrac{3}{32} \Big\lag (\pa_{\a}h) (\pa^{\a}h) -2 (\pa_{\a}h) (\pa^\a\sq w) + (\pa_\a\sq w) (\pa^\a\sq w) \Big\rag\nn
&=& \sdfrac{3}{32}\Big \lag (h-\sq w) \sq (h-\sq w)\Big\rag = 0
\eea
after integration by parts and use of the eq. of motion (\ref{scalGW}). The same substitution in the second order Ricci tensor gives
after some algebra
\be
\lag R^{(2)}_{\m\n} \rag = \Big \lag - \sdfrac{3}{32}\, \pa_{\m}(h-\sq w) \pa_{\n} (h-\sq w) + \sdfrac{1}{8} (\pa_{\m}h) (\pa_{\n}h) 
-  \sdfrac{1}{8} (\pa_{\m}\sq w) (\pa_{\n}\sq w) \Big\rag 
\label{GW2}
\ee
which is non-vanishing but also not manifestly gauge invariant. This shows that contributions from the other sets of terms in {\bf B} (or {\bf C}) 
are needed.

\subsection{Mixed terms Linear in $\vf$ and the Metric Perturbation}

The terms in the anomaly stress-energy tensor which are linear in the scalar conformalon field $\vf$ are given by
terms linear in $\vf$ in (\ref{Eab}) together with (\ref{CFterms}). Expanding these terms to first order in the
metric perturbation from flat space gives
\bea
&&T^{(1)}_{\m\n}[\vf] = -4(b+b')\,C_{(\m\ \ \n)}^{(1) \a\ \bet}\, \pa_\a \pa_\bet \vf + b' \Big\{
\tfrac{2}{3} \G^{(1)\a}_{\ \quad\m\n}\, \pa_\a \sq \vf - 4\, R^{(1)\a\!\!\!\!}\,_{(\m} \pa_{\n)}\pa_\a\vf  
+ \tfrac{8}{3}\, R^{(1)}_{\m\n}\sq \vf   \Big. \nn
&&\hspace{3cm}\left.+ \tfrac{2}{3}\, \eta_{\m\n}\left[ (\sq^2)^{(1)} \vf + 3\,R^{(1)\a\bet} \,\pa_\a\pa_\bet\vf\right]\right\}
\label{T2GW}
\eea
where (\ref{vfeomvac}) has again been used, and terms involving the $\cL_i$ terms in the anomaly stress tensor have been dropped, 
under the assumption that in free space {\it in vacuo} there are no background fields. 

The first order metric variation of the scalar $\sq^2$ operator is
\vspace{-2mm}
\be
(\sq^2)^{(1)} \vf = \tfrac{1}{2} \,(\pa_{\a} h ) (\pa^{\a} \sq \vf) + \cdots = -\tfrac{1}{2} \,h \,(\sq^2 \vf) + \cdots
\vspace{-2mm}
\ee
where the ellipsis involves total derivative terms that vanish when substituted into (\ref{T2GW}) and averaged, and
the eq.~of motion for $\vf$ (\ref{vfeomvac}) is used. Likewise since the first order metric variations of the Weyl
tensor can be expressed in terms of the first order variations of the Riemann and Ricci tensors, and these are
given in terms of $h- \sq w$ by (\ref{linEinten}), it follows again by integration by parts and repeated use of the eq. of motion
(\ref{scalGW}) that all these first order curvature terms in (\ref{T2GW}) vanish upon averaging.
Hence the only term in (\ref{T2GW}) which survives upon averaging is
\be
\lag T^{(1)}_{\m\n}[\vf]\rag = \sdfrac{2b'}{3} \Big\lag\G^{(1)\a}_{\ \quad\m\n}\, \pa_\a \sq \vf \Big\rag = 
\sdfrac{2b'}{3} \Big\lag (\pa_{\a} h^{\a\!\!}\,_{(\m}) (\pa_{\n)}\sq \vf) \Big\rag = \sdfrac{b'}{6} \Big\lag \pa_{(\m} (h+ 3 \sq w)(\pa_{\n)}\sq \vf) \Big\rag 
\ee
for scalar perturbations. Substituting for $\sq\vf$ from (\ref{scaleq}) and again freely integrating by parts under the averaging brackets
gives 
\be
\lag T^{(1)}_{\m\n}[\vf]\rag = \sdfrac{1}{8 \pi G} \,\Big\lag\sdfrac{1}{16}  \pa_{\m}(h-\sq w) \pa_{\n} (h-\sq w) + \sdfrac{1}{4} (\pa_{(\m}h) (\pa_{\n)}\sq w) 
- \sdfrac{1}{4} (\pa_{(\m}\sq w) (\pa_{\n)}\sq w) \Big\rag
\ee
which when added to (\ref{GW1}) and using (\ref{GW2}) gives the simple result
\be
T_{\m\n}^{\scriptscriptstyle SGW} =  \sdfrac{1}{8 \pi G} \, \sdfrac{1}{32} \Big\lag \pa_{\m}(h-\sq w) \pa_{\n} (h-\sq w) \Big\rag
= \sdfrac{16 \pi G b^{\prime 2}}{9} \ \Big\lag (\pa_{\m}\sq \vf)\,(\pa_{\n} \sq \vf)\Big\rag
\label{TSGW}
\vspace{-1mm}
\ee
whose gauge invariance furnishes a useful check of the calculations.

\subsection{Terms Quadratic in $\vf$}

Since the the conformalon field $\vf$ is itself first order in the metric perturbations from flat space, the $\vf$ quadratic terms in its
stress-energy tensor $T_{\m\n}[\vf]$ may be evaluated in flat space, and are given by (\ref{Eabflat})-(\ref{Cabflat}) to be
\vspace{-2mm}
\bea
&&T^{(2)}_{\m\n}[\vf] = b' \Big\lag \!\!- 2\,(\pa_{(\m}\vf) (\pa_{\n)} \sq \vf)  + 2\, (\sq\vf)(\pa_\m\pa_\n\vf)
+ \sdfrac{2}{3}\, (\pa_\a\vf)(\pa^\a\pa_\m\pa_\n\vf)
- \sdfrac{4}{3}\,( \pa_\m\pa_\a \vf)(\pa_\n\pa^\a\vf)\Big\rag\nn
&& \hspace{2cm} -\sdfrac{b'}{2\,}\, \eta_{\m\n}\, \Big\lag (\sq\vf)^2\Big\rag
= 2b'\, \Big\lag (\sq\vf) \,(\pa_{\m}\pa_{\n}\vf)\Big\rag
\label{Tphi2}
\eea
where as before integration by parts and the eq. of motion $\sq^2 \vf =0$ outside all sources have been used freely under the Brill-Hartle-Isaacson
wave averaging brackets. Substituting (\ref{sdvf})-(\ref{sdsqvf}) gives 
\be
\left(\begin{array}{c}T^{(2)}_{tt}[\vf]\\T^{(2)}_{ti}[\vf]\\T^{(2)}_{ij}[\vf]\end{array}\right)  
\rightarrow  -\frac{2b'}{r} \left(\begin{array}{c} 1\\-\br_i\\\br_i\br_j \end{array}\right) \,\Big\lag \frac{\pa}{\pa t} \left( \int d^3\bx\, J(\tilde t, \bx) \right)^2 \Big\rag = 0
\ee
for each of the components of this tensor in the far field radiation zone, when averaged over time. Thus there are no contributions to the stress-energy 
of scalar gravitational waves from the third set ({\bf C}) of terms quadratic in the $\vf$ in perturbations about flat space. These terms are associated 
instead with non-wavelike or near field effects of the anomaly sources for $\vf$. Contributions from the Weyl invariant terms (\ref{SaddW}), if present, are
also of this kind. Hence the addition of the term (\ref{SaddW}) has no effect on the stress tensor of scalar gravitational radiation.
\vspace{4mm}

The stress-energy carried by scalar gravitational waves due to the conformal anomaly is given therefore by the sum of ({\bf A}) and ({\bf B}) terms only
in (\ref{TSGW}). Note that this stress-energy is conserved and the energy density is {\it positive}:
\vspace{-3mm}
\be
T_{tt}^{\scriptscriptstyle SGW} = \sdfrac{16 \pi G b^{\prime 2}}{9} \ \Big\lag (\sq \dot \vf)^2\Big\rag \ge 0
\ee
so there is no linear instability. The outward flux from the localized sources considered in Sec. \ref{Sec:Green} is\\
\vspace{-8mm}
\bea
T_{\ i}^{t\,\scriptscriptstyle SGW} &=& \sdfrac{16 \pi G b^{\prime 2}}{9}\  \br_i \ \Big\lag (\sq \dot \vf)^2\Big\rag\nn
&=& \frac{G}{36 \pi} \frac{\br_i}{r^2} \, \Big\lag \left(\int d^3\bx \,\dot\cA (\tilde t, \bx)\right)^2\Big\rag
\eea
in the far field radiation zone. Thus for time harmonic anomaly sources (\ref{timeosc}) the power radiated by scalar gravitational
radiation per unit solid angle in the direction $\br$ is 
\be
\left(\frac{dP}{d\W}\right)_{\!_{\!\!SGW}}\hspace{-5mm}(\br)  = r^2\, \br_i \, T_{\ i}^{t\, \scriptscriptstyle SGW} 
= \frac{G\,\w^2}{72 \pi c^5} \, \big\vert \tilde A(\w|\br)\big\vert^2
= \frac{G\,\w^2}{72 \pi c^5}  \left\vert \int d^3\bx\, e^{-i \w \br \cdot \bx/c} \cA_{\w}(\bx)\right|^2
\label{SGWPower}
\ee
after time averaging, and the factors of $c$ have been re-inserted. 

Note that the multipole expansion of (\ref{SGWPower}) obtained by expanding the exponential in powers of $(\w \br \cdot \bx/c)$ 
begins with a monopole term, unlike that of transverse, tracefree gravitational waves in the classical Einstein theory, whose lowest 
order multipole is a quadrupole (\ref{TTQuad}). In the monopole approximation the total power radiated in scalar gravitational
radiation is
\be
P_{_{\!SGW}} \simeq P_{_{\!SGW}}\big\vert_{monopole}= \frac{G\w^2}{18c^5}\, \left\vert \int d^3\bx\,\cA_{\w}(\bx)\right|^2
\label{scalmonpower}
\ee
which may be compared to the lowest order multipole tensor radiation formula in Einstein's theory
\be
P_{_{TGW}} \simeq P_{_{TGW}}\big\vert_{quadrupole} = \frac{2G\w^6}{5c^9}\, \big\vert \mQ_{ij}(\w)\big\vert^2
\ee
in terms of the quadrupole moment (\ref{Quadmom}) of the classical stress tensor of the source. Thus the power radiated in scalar 
radiation is enhanced relative to that of transverse radiation by a factor of $(\w a/c)^4$,  but is suppressed by the weakness of the 
conformal anomaly stress-energy tensor source $\cA$ compared to strictly classical sources. In the next section possible astrophysical 
sources of scalar gravitational waves are considered and the amplitude and power radiated for these sources estimated.

\section{Astrophysical Sources of Scalar Gravitational Waves}
 \label{Sec:Astro}
\vspace{-1mm}
The typical curvature invariant in the vicinity of a completely collapsed star is of order
\vspace{-2mm}
\be
R_{\m\n\a\bet}R^{\m\n\a\bet} = \frac{48 (GM)^2}{r^6} \le \frac{3}{4(GM)^4}
\vspace{-2mm}
\ee
at the star's Schwarzschild radius. This corresponds to a very small energy density of
\vspace{-2mm}
\be
\rho_{_{\!R^2}}= 5 \times 10^{-38}\, \left(\frac{M_\odot}{M}\right)^{\!4} {\rm erg/cm}^3\,.
\vspace{-2mm}
\ee
Substituting this source of scalar gravitational waves into (\ref{scalpert}) gives gravitational potentials 
\be
\Y_A = \Y_C \simeq - \frac{Gb' }{3r}\, \frac{4 \pi (2GM)^3}{3}\, \frac{3}{4(GM)^4}
= -\frac{8\pi b'}{3 M r} \simeq 10^{-91}\left(\frac{ |b'|}{\hbar}\right) \left(\frac{M_{\odot}}{M}\right) \left(\frac{\rm kpc}{r}\right)
\label{gwampR}
\ee
far below any possibility of direct detection. The reason for such enormous suppression
from curvature sources is essentially the same factor of $\hbar G R \sim (M_{Pl}/M)^2 $ in the quantum
anomaly effective action relative to the classical Einstein-Hilbert action as that which appears in local EFT
treatments. For this reason also the effects of the curvature squared terms in the anomaly EFT are
far too small to be observed in weak gravitational fields, such as that of the existing binary or double
pulsar tests of GR. As a consequence the anomaly EFT (\ref{Seff}) easily passes these observational tests.

The possibility of much larger effects arise only when one considers the non-curvature gauge field sources 
in the trace anomaly. The electromagnetic trace anomaly becomes relevant above the two-electron mass-energy 
threshold $2m_e c^2 \simeq 1.02$ MeV, and the QCD anomaly at least above the light $u$ and $d$ quark mass-energy 
thresholds of approximately $10$\,MeV. These require very high energy astrophysical environments, but energy scales 
still very far below the Planck energy scale, where the EFT defined by (\ref{Seff}) should be reliable.

In the QED case, highly magnetized neutron stars (`magnetars') are believed to have magnetic fields up to $10^{15}$ Gauss,
which exceeds the electrodynamic critical field of $B_c \simeq 4 \times 10^{13}$ Gauss corresponding to $2m_e c^2$ \cite{mag}.
Taking into account the coefficient of the trace anomaly in QED, the magnetar field provides a source of scalar gravitational
waves of strength
\be
\cA_{mag} =  -\frac{e^2}{24 \pi^2} F_{\m\n}F^{\m\n} = -\frac{\a B^2\!}{\!3 \pi}
 \simeq  8 \times 10^{26} \left(\frac{B}{10^{15}\,\rm Gauss}\right)^{\!2}\rm erg/cm^3
\label{Amag}
\ee
Since a typical neutron star radius is $12$ km, the maximum volume over which this energy density applies is of order
$7 \times 10^{18}$ cm$^3$, giving a total magnetic field energy of order of $6 \times 10^{45}$ ergs.
The scalar gravitational wave produced at a distance $r$ from such a highly magnetized source
has magnitude
\be
\Y_A= \Y_C \simeq -\frac{G}{3rc^4} \int d^3 \bx\, \cA_{mag} \lesssim 5 \times 10^{-26} \left(\frac{\rm kpc}{r}\right)\,.
\label{scalmag}
\ee
The estimate (\ref{scalmag}) is many orders of magnitude greater
than (\ref{gwampR}) but still several orders of magnitude below the sensitivity of present gravitational wave detectors. 
When the electron mass is not neglected, because of decoupling of the triangle diagram responsible for the QED trace
anomaly, $\cA_{mag}$ is suppressed by a factor of order $(\hbar \w/m_e c^2)^2$ where $\w$ is a characteristic frequency
of time variation of the electromagnetic field strength \cite{GiaMot}. Thus (\ref{scalmag}), small as it is, will be suppressed
further as a source of scalar gravitational radiation for more slowly varying magnetic fields, for weaker fields, or for
strong fields exceeding the critical field but extending only over smaller volumes.

This suppression due to decoupling of the anomaly at low frequencies is also the reason why the power radiated
in scalar gravitational waves is negligible compared to classical quadrupolar radiation in binary pulsar systems,
such as the Hulse-Taylor binary and double pulsar. For the double pulsar the orbital period of $2.454\,$ hours
gives an angular frequency $\w \simeq 7.11 \times 10^{-4}\,$ sec \cite{dblpul}. This gives a suppression factor in the anomaly
source $\cA$ of $(\hbar \w/m_e c^2)^2 \simeq 2.10 \times 10^{-49}$, which enters the scalar power radiated (\ref{scalmonpower})
squared. Taking all the factors in (\ref{scalmonpower}) into account gives a total power emitted in scalar radiation of order
\be
\left(\frac{dE}{dt}\right)_{\!\!scalar} \sim \frac{1}{8\, (4\pi b' )} \left[ \int \!d^3 \bx \left(\frac{\a B^2}{3 \pi}\right) \right]^2
\left(\frac{\hbar\, \w}{m_e c^2}\right)^{\!4} \sim 1.1 \times 10^{12}\, \rm erg/sec
\label{radblpulsar}
\ee
where a magnetic field of $2 \times 10^{12}\,$ Gauss was assumed and the value of $b'$ from (\ref{bprime}) was
used with $N_v = 1$ for the electromagnetic field. In principle the energy radiated in scalar gravitational waves will
cause the orbit of the double pulsar to decay and its orbital period to decrease, above the decrease predicted
due to quadrupolar radiation in Einstein's classical theory. However the comparable estimate for classical transverse
tensor gravitational radiation from the same double pulsar system is of order $4 \times 10^{32}\,$ erg/sec. Thus the
effect of the additional scalar radiation is smaller than one part in $10^{20}$ in this system, and completely negligible.
The effect in the Hulse-Taylor binary pulsar is even smaller. Although a more accurate analysis is called for,
these rough estimates indicate that the anomaly EFT will easily pass all existing pulsar tests from electromagnetic sources

The most promising source for generating scalar gravitational waves in significant and potentially detectable quantities is the
$SU(3)_{color}$ trace anomaly
\be
\cA_{QCD} =  (11 N_c - 2 N_f)\,\frac{\a_s}{24 \pi}  \, G^a_{\m\n}G^{a\m\n} \simeq -4.8 \times 10^{36}\, \rm erg/cm^3
\label{QCDanom}
\ee
for $N_c =3$ colors and $N_f = 2$ light fermion species, where the last value is that of an effective bag `constant'
of $\rho_{bag}= -p_{bag} \simeq 750$ Mev/fm$^3$ in nuclear matter \cite{Shur}. The chemical potential
dependence of the bag `constant' of dense nuclear matter has been estimated, with values of the QCD trace
anomaly similar to or even higher than (\ref{QCDanom}) possible at baryon chemical potentials of $\m \simeq 1.6$\, GeV
thought to exist in neutron star cores \cite{Fra}.  This energy density is more than $10$ orders of magnitude larger than
the QED value in a strong magnetar field (\ref{Amag}), and thus capable in principle of producing scalar gravitational
waves $10$ orders of magnitude larger than (\ref{scalmag}), potentially within the detectable range.

As with the QED source, an important caveat for this estimate is that the QCD gauge field anomaly proportional to $N_f=2$ is a
source for the scalar gravitational waves only to the extent that light $u$ and $d$ quark masses of $5$ to $10$ MeV can be
neglected. This corresponds to a nuclear time scale of $7 \times 10^{-23}\,$ sec., or nuclear distance scales of $20$ Fermi.
Gluonic vacuum fluctuations responsible for the $N_c=3$ in (\ref{QCDanom}) are presumably likewise suppressed
on distance and time scales larger than the confinement scale of a few Fermi. For lower energy processes, neither the
light quarks nor the gluons can be treated as massless fields, and the QCD trace anomaly $\cA_{QCD}$ will be suppressed
by a factor of order $(\hbar \w/m_{\pi} c^2)^2$, by non-perturbative effects of confinement. At these lower energies
and frequencies, one should use the low energy meson EFT of the strong interactions, rather than the QCD anomaly.

The most vigorous disturbance of the QCD gluonic vacuum which can excite this astrophysical source of the conformal anomaly is
in the initial formation of the neutron star (NS), or its collision, coalescence and merger with another compact stellar object. In such
processes density estimates such as (\ref{QCDanom}) may be applicable. Thus the scalar gravitational wave amplitude from neutron
star formation or binary coalescence is estimated to be
\be
\Y_A= \Y_C \simeq -\frac{G}{3rc^4} \int d^3 \bx\, \cA_{NS} \, \eta \simeq  3 \times 10^{-21} \left(\frac{100\,\rm Mpc}{r}\right) \, \eta
\label{scalQCD}
\ee
where $\eta <1$ is volume fraction of the neutron star in which the gluonic condensate is excited by high energy
interactions above the QCD deconfinement threshold ($\sim 165\,$ MeV) in the formation or coalescence event.

Clearly more careful estimates are needed, and will require detailed modeling of the time dependence of the nuclear constituents 
and gluonic condensate in realistic neutron star formation and merger events. Still, the signal generated by a time dependent QCD 
trace anomaly by nuclear matter in a NS formation or merger event is almost certainly the strongest astrophysical source of scalar 
gravitational waves, potentially detectable by present or planned gravitational wave detectors, meriting such a detailed
study. If NS binary coalescence events are the sources of detectable scalar gravitational radiation, this emission
should follow closely upon transverse, traceless GW emission from binary inspiral, suggesting the time co-incidence
study of production of both GW polarizarization states by the same event(s) may be a promising observational strategy.
Conversely, non-detection of scalar gravitational waves from NS or black hole candidate coalescence events
could provide potentially interesting constraints on nuclear equations of state and/or the mechanisms of gravitational
wave generation by the QCD trace anomaly.

\section{Summary and Outlook}
\label{Sec:Conc}

In this paper the extension and modification of Einstein's theory of classical General Relativity by the infrared relevant
effects of the conformal trace anomaly have been analyzed in the near flat space limit. The effective action (\ref{Seff})
due to the quantum effects of the anomaly defines the low energy EFT of gravity with an additional scalar field.
The scalar field $\vf$ is called a {\it conformalon} because it is closely related to and mixes with the conformal
mode of the usual spacetime metric, making it dynamical. Although reminiscent of the scalar-tensor theory introduced by
Fierz, Jordan, Brans and Dicke \cite{BrDicke}, (\ref{Seff}) is quite different in several important respects, foremost among them that
it its exact form is dictated by the known quantum effect of the conformal anomaly in QFT. As such the EFT presented
here does not fall into the class of scalar-tensor theories or effective field theories of scalars or modified gravity theories
usually considered \cite{Will}. Most notably, the scalar $\vf$ does not couple directly to the trace of the matter stress tensor,
but only to higher order curvature invariants $E,C^2$ of (\ref{EFdef}), or the gauge field scalars such as $F_{\m\n}F^{\m\n}$ or
$G^a_{\m\n}G^{a\,\m\n}$. This feature of the scalar conformalon most clearly distinguishing it from other possible scalar
dilaton-like degrees of freedom or other modified gravity theories is a result of the Wess-Zumino consistency relation (\ref{SanomWZ})
derived from the conformal anomaly (\ref{WZfour})-(\ref{WZcons}), which strictly constrains the form of the effective action.

The equations of the resulting semi-classical EFT are Einstein's eqs.~(\ref{Ein}) with an additional covariantly
conserved source $T_{\m\n}[\vf]$, (\ref{Ttens})-(\ref{CFterms}) of the scalar conformalon field describing
certain quantum vacuum effects of the trace anomaly, supplemented by the eq.~of motion (\ref{phieom}) for
this field itself. Linearized around flat spacetime with $\vf = 0$, the EFT describes scalar gravitational waves, in addition to
the tensorial transverse, traceless waves of the linearized Einstein theory. Although the anomaly EFT has a fourth order
differential operator and $\vf$ solves $\sq^2 \vf =0$, {\it only the half of these solutions} with $\sq \vf \neq 0$
couple to the gauge invariant scalar gravitational metric perturbations through (\ref{UpsACsoln}).
The other half of the solutions to (\ref{phieom}) satisfying $\sq \vf =0$ are a passive fixed background which
are not coupled to the metric and do not contribute to scalar gravitational radiation at linearized order.
This is a very important consideration relevant to the stability of the EFT and any difficulties
encountered with local higher derivative theories. In particular there is no linearized instability of flat space
in the anomaly EFT, requiring the discarding of high frequency or `runaway' solutions as there is in higher
derivative local theories of gravity with $R^2$ or $C^2$ terms \cite{Simon,AndMolMot}. 

Since the anomaly effective action is quadratic in $\vf$, the excluded half of the solutions of the fourth order
linear eq.~of motion (\ref{phieom}) cannot couple to the spacetime metric except through the non-linear terms
in the usual Einstein-Hilbert action of General Relativity itself, through their mutual gravitational interaction. 
These interactions are naturally suppressed by the weakness of the gravitational coupling $G$, which must 
come together with some energy scale squared for dimensional reasons. Since around flat spacetime
the only energy scale available is that of the perturbations, one must expect the other half of the solutions 
of the fourth order anomaly EFT to couple only when their energy scale approaches the Planck energy of 
$10^{19}$ GeV, by which point the EFT approach to gravity has clearly broken down. In curved spacetime 
backgrounds possessing additional length scales or horizons, the breakdown of the EFT approach 
may occur at lower energy scales, such as in the vicinity of horizons \cite{MotVau,DSAnom}. Issues 
arising from fourth order anomaly effective action (\ref{Sanom}) including the quantization and Hamiltonian of the 
conformalon field will be taken up in a separate publication, inasmuch as they do not affect the 
conclusions of scalar gravitational waves with classical gauge invariant potentials obeying the {\it second order} 
eq.~(\ref{scalpotwvs}). Clearly the full treatment of the anomaly quantum effective theory is required to settle all 
questions of the fourth order theory in either its local or non-local form.

When localized sources for the scalar gravitational waves are considered, the scalar metric perturbations
have a monopole form, falling off with distance from the source as $1/r$ according to (\ref{scalpert}), or
(\ref{scalgwave}) for time harmonic sources. The corresponding linearized Riemann tensor components for
scalar gravitational waves needed for the response of test masses in GW detectors are given by eqs. (\ref{Riescal}).
The energy flux and power emitted per unit solid angle in scalar gravitational waves are given in the
EFT by (\ref{SGWPower}), or (\ref{scalmonpower}) in the monopole approximation.

Although all possible terms in the trace anomaly $\cA$ can act as sources for the scalar conformalon field
and therefore scalar gravitational radiation, the higher derivative curvature sources are far too small to
contribute appreciably. The electromagnetic trace anomaly of QED is much larger but still very likely
too weak to produce observable scalar GWs or indirect effects, such as energy loss from even the most
highly magnetized objects known. Thus the EFT associated with the conformal anomaly easily passes
all present observational tests from the binary and double pulsar systems.

The most promising non-negligible sources of scalar gravitational waves in the EFT are those due to the QCD
trace anomaly, excited in neutron star formation, or binary coalescence with another NS or other collapsed star.
In such systems the rough preliminary estimate (\ref{scalQCD}) indicates that scalar gravitational radiation may
be produced with large enough amplitude to be observed by present and future planned GW detectors as
a burst event. The power radiated in scalar gravitational waves from the QCD anomaly is  potentially many
orders of magnitude larger than (\ref{radblpulsar}) from strong magnetic sources, but as proportional to
a factor $\eta^2 < 1$ subject to considerable uncertainty. A more careful quantitative calculation of the sources of
scalar GWs in NS merger events is clearly needed, with the development of detailed models of the
scalar waveforms expected in the EFT of gravity as now becoming available in standard GR \cite{CorSYunPre}.
Even at this preliminary stage an open search for scalar `breather mode' gravitational waves by present
and future detectors is indicated \cite{MagNic}, with a search strategy for burst events coincident with inspirals
producing transverse, traceless GWs seeming to be the most promising approach.

Energy densities as large as (\ref{QCDanom}) from strong interactions and the QCD trace anomaly also
imply that the non-linear effects on the geometry of the anomaly stress-energy tensor, neglected in the
present analysis around flat space, should be taken into account in a fully consistent treatment. These
non-linear effects of the anomaly stress-energy may be comparable in importance to classical GR effects
in the final stages of NS inspiral and coalescence with a second compact object, either itself another
NS or a black hole/gravastar candidate \cite{PNAS}. The gravastar alternative to black holes
may also be a source of scalar radiation through its interior scalar condensate being partly composed
of and coupled to gluonic degrees of freedom in QCD.

In the early universe when temperatures and energies greater than $165\,$MeV were reached, the QCD
anomaly was fully unsuppressed. Thus these epochs up to the QCD phase transition, or at even higher energies
the electroweak phase transition are possible sources of a cosmological stochastic background of scalar
gravitational radiation, albeit in lower frequency ranges, potentially detectable by the next generation
of space-based gravitational wave detectors, such as LISA.

\vfill\eject

\centerline{\it Acknowledgements}
\vspace{1mm}
The author gratefully acknowledges conversations with R. Vaulin at the initial stages of this work, and P. R. Anderson for
careful reading of the manuscript. This work was supported in part by a grant from the Los Alamos National Laboratory
NSEC Center for Space and Earth Science (CSES).

\end{document}